\documentclass{mn2e} 
\usepackage[dvips]{graphicx} 
\usepackage{amssymb} 
\usepackage{amsbsy} 
\usepackage{txfonts} 
\newcommand{\apj}{{ApJ}} 
\newcommand{\mnras}{{MNRAS}}

\newcommand{\ledd}{L_{{\rm Edd}}} 
\newcommand{\medd}{{\dot{M}_{\rm Edd}}} 
\title[]{Gamma-ray emission from proton-proton interactions in hot accretion flows}   
\author[A. Nied\'zwiecki, F.-G.\ Xie, A. St\c epnik]  
{Andrzej Nied\'zwiecki,$^1$\thanks{E-mail: niedzwiecki@uni.lodz.pl (AN),
fgxie@shao.ac.cn (FGX), agastepnik82@gmail.com (AS)} 
Fu-Guo Xie$^{2,3}$\footnotemark[1]
and Agnieszka St\c epnik$^{1}$\footnotemark[1]\\  
$^1$Department of Astrophysics, University of \L \'od\'z, Pomorska 149/153, 
90-236 \L \'od\'z, Poland\\  
$^2$Key Laboratory for Research in Galaxies and Cosmology, Shanghai
  Astronomical Observatory, Chinese Academy of Sciences,\\  
80 Nandan Road, Shanghai 200030, China\\  
$^3$Kavli Institute for Astronomy and Astrophysics, Peking University, 
Beijing 100871, China\\  
}

\pagerange{\pageref{firstpage}--\pageref{lastpage}} 
\pubyear{2012} 
\begin{document} 
\maketitle 
\label{firstpage} 
 
\begin{abstract}

We present a model of $\gamma$-ray emission through neutral pion production and decay in two-temperature accretion flows around supermassive black holes. We refine previous studies of such a hadronic $\gamma$-ray emission by taking into account (1) relativistic effects in the photon transfer and (2) absorption of $\gamma$-ray photons in the radiation field of the flow. We use a fully general relativistic description of both the radiative and hydrodynamic processes, which allows us to study the dependence on the black hole spin. The spin value strongly affects the $\gamma$-ray emissivity within $\sim 10$ gravitational radii. The central regions of flows with the total luminosities $L \la 10^{-3}$ of the Eddington luminosity   ($L_{\rm Edd}$) are mostly transparent to photons with energies below 10 GeV, permitting investigation of the effects of space-time metric. For such $L$, an observational upper limit on the $\gamma$-ray (0.1 -- 10 GeV) to X-ray (2 -- 10 keV) luminosity ratio of $L_{\rm 0.1-10 GeV}/L_{\rm 2-10 keV} \ll 0.1$ can rule out rapid rotation of the black hole; on the other hand, a measurement of  $L_{\rm 0.1-10 GeV}/L_{\rm 2-10 keV} \sim 0.1$ cannot be regarded as the evidence of rapid rotation, as such a ratio can also result from a flat radial profile of $\gamma$-ray emissivity (which would occur for nonthermal acceleration of protons in the whole body of the flow). At  $L \ga 10^{-2} L_{\rm Edd}$, the $\gamma$-ray emission from the innermost region is strongly absorbed and the observed $\gamma$-rays do not carry information on the value of $a$. We note  that if the X-ray emission observed in Centaurus A comes from an accretion flow, the hadronic $\gamma$-ray emission from the flow should contribute significantly to the MeV/GeV emission observed from the core of this object, unless it contains a slowly rotating black hole and protons in the flow are thermal.

\end{abstract}
\begin{keywords} 
accretion, accretion discs -- black hole physics -- gamma-rays: theory
\end{keywords} 
 
\section{Introduction} 
\label{intro} 

Early investigations of  black hole accretion flows indicated that tenuous flows can develop a two-temperature structure, with proton temperature sufficient to produce a significant $\gamma$-ray luminosity above 10 MeV through $\pi^0$ production (e.g.\ Dahlbacka, Chapline \& Weaver 1974). The two-temperature structure is an essential feature of the optically-thin, advection dominated accretion flow (ADAF) model, which has been extensively studied and   successfully applied to a variety of black hole systems (see, e.g., reviews in Yuan 2007, Narayan \& McClintock 2008, Yuan \& Narayan 2013) over the past two decades, following the work of Narayan \& Yi (1994). Mahadevan, Narayan \& Krolik (1997; hereafter M97) pointed out that $\gamma$-ray emission resulting from proton-proton collisions in ADAFs may be a signature allowing to test their fundamental nature. The model of M97 relied on a non-relativistic ADAF model and their computations were improved by Oka \& Manmoto (2003; hereafter OM03) who used a fully general relativistic (GR) model of the flow. However, both M97 and OM03 neglected the Doppler and gravitational shifts of energy as well as gravitational focusing and capturing by the black hole, which is a major deficiency because the $\gamma$-ray emission is produced very close to the black hole's horizon. Furthermore, both works neglected the internal absorption of $\gamma$-ray photons to pair creation, which effect should be important in more luminous systems. 

ADAFs are supposed to power low-luminosity AGNs, like Fanaroff-Riley  type  I (FR I) radio galaxies or low-luminosity Seyfert galaxies, and a measurement, or even upper limits on their $\gamma$-ray emission, may  put interesting constraints on the properties of the source of high-energy radiation in such objects. M97 and OM03 considered only the {\it CGRO}/EGRET source in  the direction of the Galactic Center for such an analysis. Significant progress in exploration of the $\gamma$-ray activity of AGNs which has been made after their works, thanks to the {\it Fermi} mission, motivates us to develop a more accurate model of the hadronic $\gamma$-ray emission from ADAFs. Detections of $\gamma$-ray emission from objects with misaligned jets (e.g.\ Abdo et al.\ 2010b) are most relevant for our study. Their $\gamma$-ray radiation is usually explained as a jet emission; we show that emission from an accretion flow may be a reasonable alternative, at least in some FR Is.  We focus on modelling of radiation in 100 MeV -- 10 GeV energy range, relevant for the {\it Fermi}-LAT measurements of the FR I radio galaxies (Abdo et al.\ 2010b) and over  which the upper limits in Seyfert galaxies are derived (Ackermann et al.\ 2012). 

The dependence of the $\gamma$-ray luminosity on the black hole spin parameter makes a particularly interesting context for such an investigation. Already a rough estimate by Shapiro, Lightman \& Eardley (1976) indicated a strong dependence of the $\gamma$-ray luminosity from a two-temperature flow on the spin of the black hole and, then, they suggested that this effect may serve as a means to measure the spin value (see also Eilek \& Kafatos 1983 and Colpi, Maraschi \& Treves 1986). OM03, who made GR calculations for the modern ADAF model, found a dramatic dependence of the $\gamma$-ray luminosity on the spin value in models with thermal distribution of proton energies, however, they concluded that the dependence is weak if protons have a nonthermal distribution. In this work we extend the analysis of this issue and clarify some related properties.

We find global solutions of the hydrodynamical ADAF model, which follows Manmoto (2000), and use them to compute the $\gamma$-ray emission. Similarly to M97 and OM03 we take into account emission resulting from thermal and nonthermal distribution of proton energies; we use similar phenomenological models, with some modifications which allow to illustrate separately effects due to local distribution of proton energies and to radial profile of $\gamma$-ray emissivity. We also use our recently developed  model of global Comptonization (Nied\'zwiecki, Xie \& Zdziarski 2012; hereafter N12, see also Xie et al.\ 2010) to  compute the X-ray emission, which allows to investigate the internal absorption of $\gamma$-ray photons to pair creation in the flow. 

In our computations we assume a rather weak magnetic field, with the magnetic pressure of 1/10th of the total pressure, supported by results of the magnetohydrodynamic (MHD) simulations in which amplification of magnetic fields by the magneto-rotational instability typically saturates at such a ratio of the magnetic to the total pressure (e.g.\ Machida, Nakamura \& Matsumoto 2004, Hirose et al.\ 2004, Hawley \& Krolik 2001). We investigate the dependence on the poorly understood parameter in ADAF theory, $\delta$, describing the fraction of the turbulent dissipation that directly heats electrons in the flow. We take into account only one value of the accretion rate, but the considered ranges of the spin and $\delta$ parameters yield a rather large range of bolometric luminosities of $\sim 10^{-4}$ to $10^{-2}$ of the Eddington luminosity. In our paper we present both the spectra affected by $\gamma \gamma$ absorption and those neglecting the absorption effect; the latter may be easily scaled to smaller accretion rates, for which the $\gamma \gamma$ absorption becomes unimportant.


\begin{figure*} 
\centerline{
\includegraphics[height=5cm]{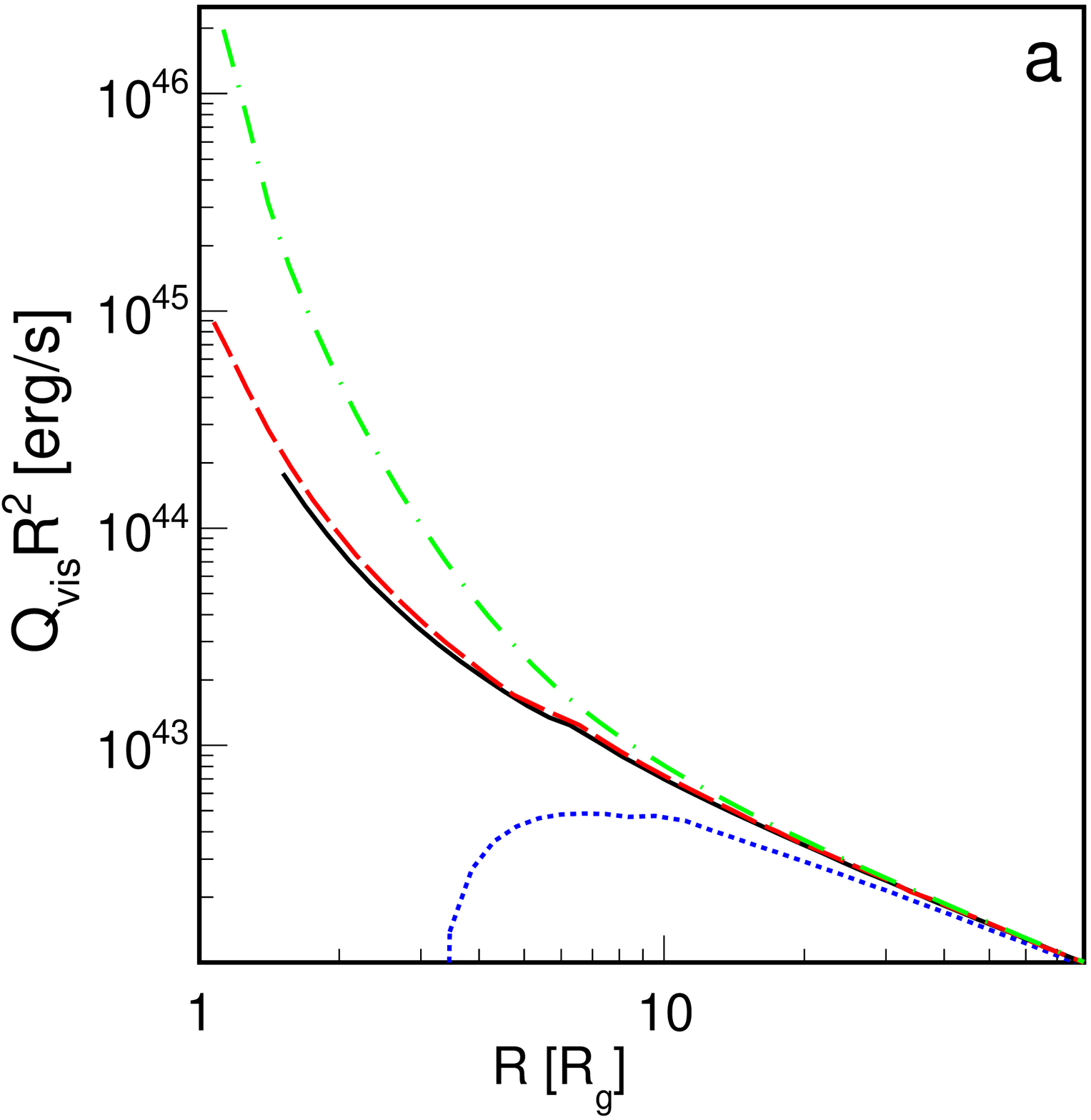}\hspace{1pc}\includegraphics[height=5cm]{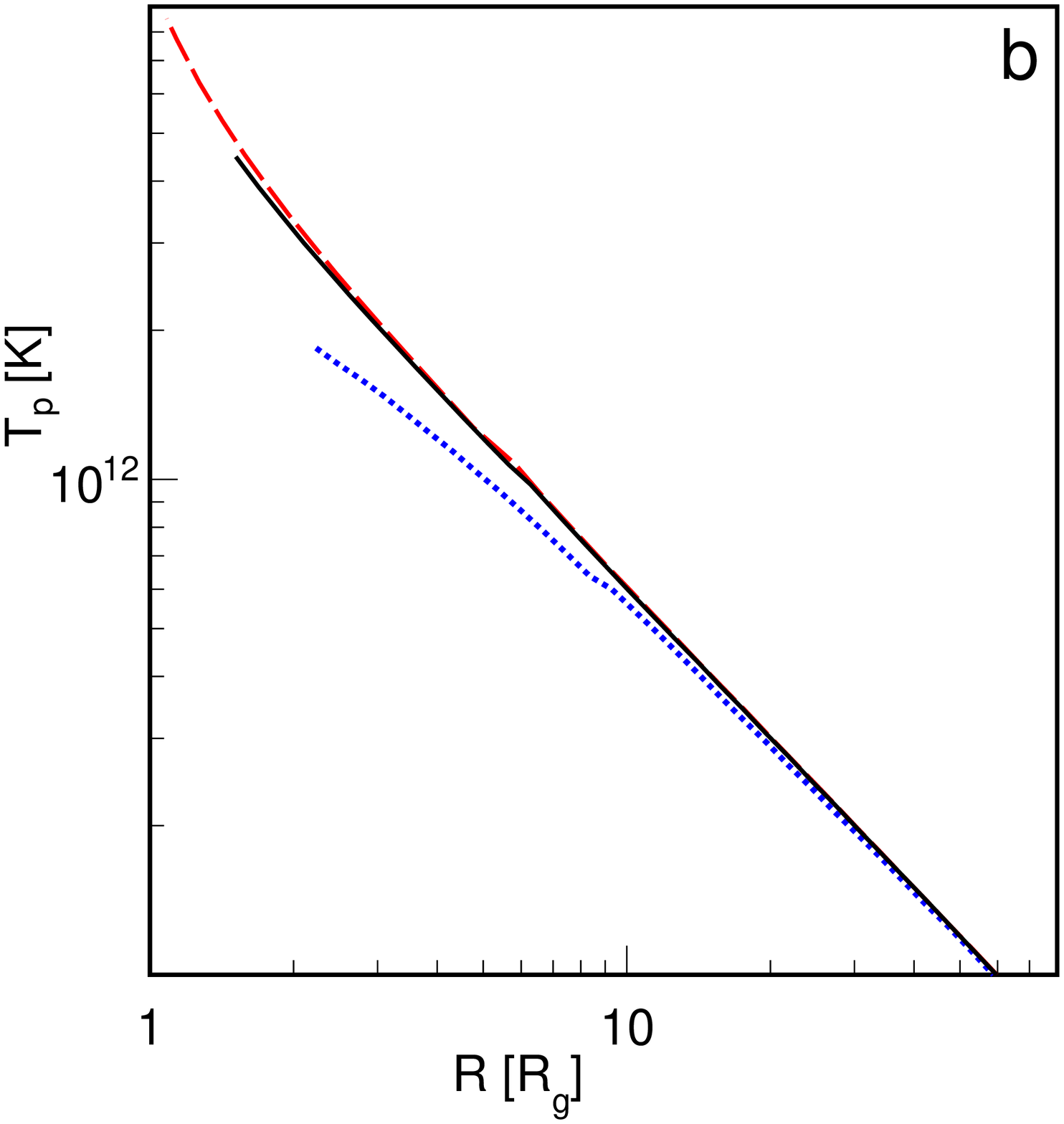}\hspace{1pc}
\includegraphics[height=5cm]{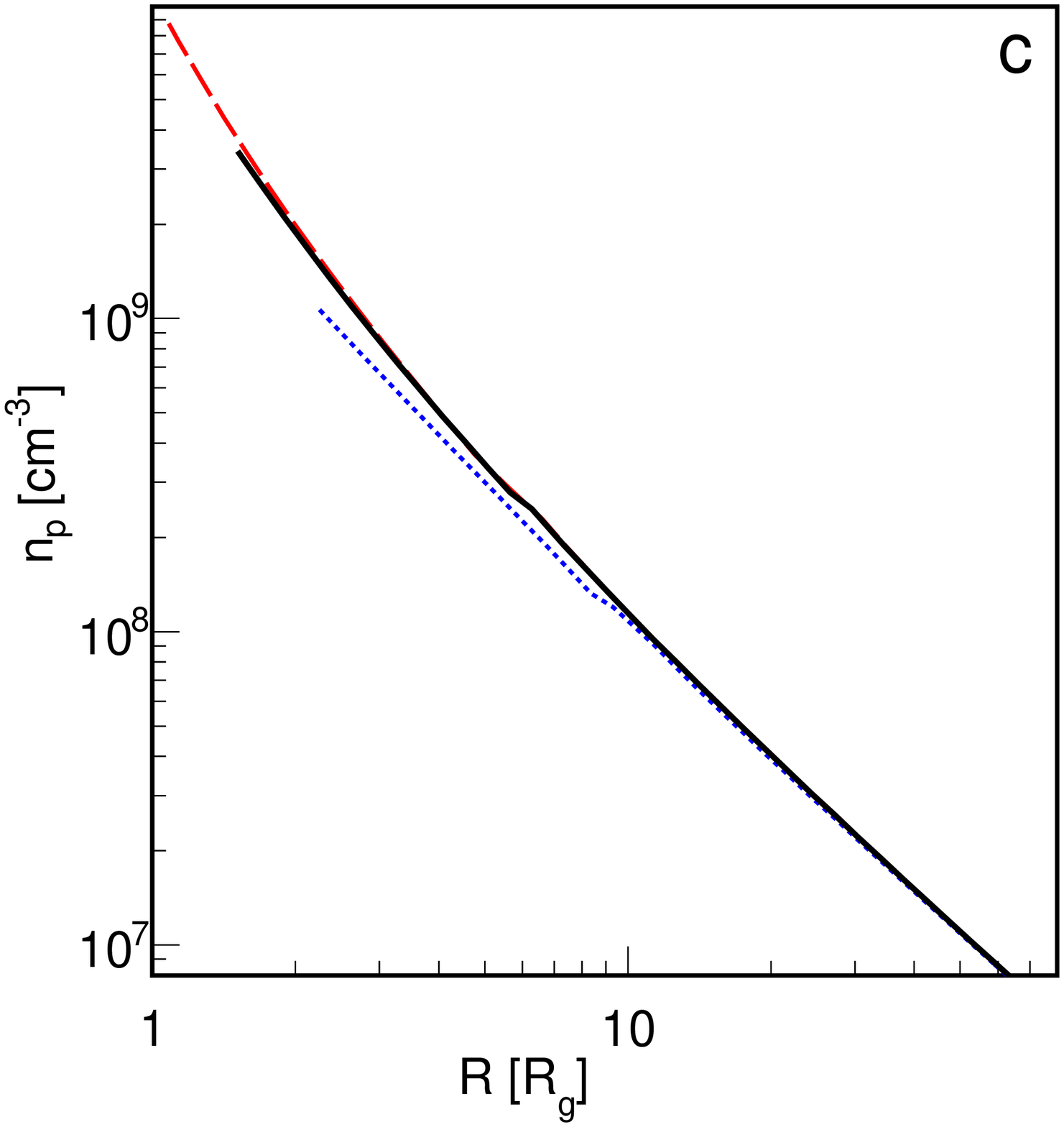}}
\caption{Radial profiles of the dissipative heating rates, $Q_{\rm vis}$ (a), the proton temperature (b) and the proton number density (c) of our hot-flow solutions for $\delta=10^{-3}$. In all panels, the dashed (red) lines are for $a=0.998$, the solid (black) lines are for $a=0.95$ and the dotted (blue) lines are for $a=0$. In panel (a), $Q_{\rm vis}$ denotes a vertically  integrated rate, so $Q_{\rm vis}R^2$ gives the heating rate (per unit  volume) times volume. The green (dashed) line in panel (a) shows $Q_{\rm vis}$ in the superhot solution (see text) for $a=0.998$. $M =2 \times 10^8 M_{\odot}$, $\dot m=0.1$, $\alpha=0.3$ and $\beta_{\rm B} = 0.9$ in this and all further figures in this paper.
}
\label{fig:1} 
\end{figure*}

\section{Hot flow model}
\label{sec:flow}

We consider a black hole, characterised by its mass, $M$, and angular momentum, $J$, surrounded by a geometrically thick accretion flow with an accretion rate, $\dot M$. We define the following dimensionless parameters: $r = R / R_{\rm g}$, $a = J / (c R_{\rm g} M)$, $\dot m = \dot M / \medd$, where $\medd= \ledd/c^2$, $R_{\rm g}=GM/c^2$ is the gravitational radius and $\ledd \equiv 4\pi GM m_{\rm p} c/\sigma_{\rm T}$ is the Eddington luminosity. Most results presented in this work correspond to $M =2 \times 10^8 M_{\odot}$, in Fig \ref{fig:6}a we present also results for $M =2 \times 10^6 M_{\odot}$.  We consider $\dot m=0.1$ and three values of the spin parameter, $a=0$, 0.95 and 0.998. The inclination angle of the line of sight to the symmetry axis is given by $\theta_{\rm obs}$. We assume that  the density distribution is given by $\rho(R,z)=\rho(R,0) \exp(-z^2/2H^2)$, where $H$ is the scale height at $r$. We assume the viscosity parameter of $\alpha=0.3$ and the ratio of the gas pressure (electron and ion) to the total pressure of $\beta_{\rm B} = 0.9$. The fraction of the dissipated energy which heats directly electrons is denoted by $\delta$. 

Our calculations of hadronic processes are based on global solutions  of the fully GR hydrodynamical model of two-temperature ADAFs, described in N12, which follows closely the model of Manmoto (2000). Here we recall only the ion energy equation, which is most important for the present study: 
\begin{equation} 
0 = (1 - \delta)Q_{\rm vis}  + Q_{\rm compr} - \Lambda_{\rm ie} - Q_{\rm int},
\label{eq:ions} 
\end{equation}
where $\Lambda_{\rm ie}$ is the Coulomb rate, the compressive heating and the advection of the internal energy of ions, respectively, are given by \begin{equation} 
Q_{\rm compr} = - {\dot M p_{\rm i} \over 2 \pi R \rho} 
{{\rm d}\ln \rho \over {\rm d}R},~~~
Q_{\rm int} = - {\dot M p_{\rm i} \over 2 \pi R \rho (\Gamma_{\rm i} - 1)} 
{{\rm d}\ln T_{\rm i} \over {\rm d} R}, 
\end{equation}
and the viscous dissipation rate, per unit area, is given by  
\begin{equation} 
Q_{\rm vis} = -  \alpha p  H (2 \upi)^{1/2}  \frac{\gamma_\phi^4A^2}{r^7} \frac{{\rm d}\Omega}{{\rm d}r}, 
\label{eq:qdiss} 
\end{equation}
where $p =  (p_{\rm i} + p_{\rm e})/\beta_{\rm B}$, $p_{\rm i}$  is the ion pressure, $p_{\rm e}$ is the electron pressure, $\Gamma_{\rm i}$ is the ion adiabatic index, $\Omega$ is the angular velocity of the flow, $\gamma_\phi$ is the Lorentz factor of the azimuthal motion and $A=r^4 + r^2a^2 + 2ra$. The form of the energy equation given in equation (\ref{eq:ions}) is standard in ADAFs theory, although actually it should include an additional term describing the direct cooling of protons to pion production, $Q_{\gamma}$. In our calculation of hadronic processes we find that $Q_{\gamma}$ is approximately equal to $\Lambda_{\rm ie}$ at $r<10$. At $\dot m = 0.1$, considered in this work, both $Q_{\gamma}$ and $\Lambda_{\rm ie}$ are much smaller, by over 3 orders of magnitude, than the effective heating $Q_{\rm vis}  + Q_{\rm compr}$ and the heating is fully balanced by the advective term, $Q_{\rm int}$. This justifies our neglect of the direct hadronic cooling.

The only  difference between our GR model and that of Manmoto (2000) involves the simplifying assumption of ${\rm d~ln}(R)/{\rm d~ln}(H)=1$ adopted in the latter; we do not follow this simplification and an exact $H(R)$ profile is considered in all our hydrodynamic equations. We note that the simplification has a considerable effect in the central part of the flow, e.g.\ it results in an underestimation of the proton temperature by a factor of $\sim 1.5$ within the innermost $10 R_{\rm g}$. Applying the above simplifying assumption we get exactly the same flow parameters as Manmoto (2000); note, however, that Manmoto (2000) assumed an equipartition between the gas and plasma pressures, with $\beta_{\rm B}=0.5$, which in general gives a smaller proton temperature than $\beta_{\rm B}=0.9$ assumed here. In particular, for $a=0$ and $\delta=10^{-3}$, models with $\beta_{\rm B}=0.9$ give the proton temperature larger by a factor of $\simeq 4$, close to the horizon, than models with $\beta_{\rm B}=0.5$. This underlies also the differences in the $\gamma$-ray luminosity levels between the thermal models of Oka \& Manmoto (2003) and ours, as discussed in Section 3.

To obtain global transonic solutions we have to adjust the specific angular momentum per unit mass accreted by the black hole, for which the accretion flow passes smoothly through the sonic point, $r_{\rm s}$. We note that this condition permits for two kinds of solutions, below referred to as a 'standard' and a 'superhot' solution. The latter (superhot) has much larger proton temperature and density, furthermore, the sound speed is large and the sonic point located in the immediate vicinity of the event horizon, e.g.\ $r_{\rm s} \simeq 1.2$ for $a=0.998$. In the standard solutions the sonic point is located at larger distances, $r_{\rm s} > 2$. Taking into account rather extreme properties of the superhot solutions (specifically, a very large magnitude of $Q_{\rm vis}$ illustrated in Fig.\ \ref{fig:1}(a) and discussed in Section \ref{sec:diss}) we neglect them in this work and for all values of $a$ we consider only the standard solutions which are consistent with solutions of the model investigated in several previous studies (e.g.\ Manmoto 2000, Yuan et al\ 2009, Li et al.\ 2009). Note, however, that in our previous works (N12, Nied\'zwiecki, Xie \& Beckmann 2012) we considered the superhot solution with $a=0.998$, then, the results for $a=0.998$ discussed in those works correspond to flows with larger proton temperature and density (both by a factor of $\sim 5$) than these considered in the present study.

Fig.\ \ref{fig:1} shows the dependence on the black hole spin of some parameters from our solutions which are crucial for the hadronic $\gamma$-ray production. Rotation of the black hole stabilizes the circular motion of the flow which yields a higher density (through the continuity equation). Furthermore, the stabilized rotation of the flow results in a stronger dissipative heating giving a larger proton temperature for larger $a$. All these differences are significant only  within the innermost $\sim 10 R_{\rm g}$.

\begin{figure*} 
\centerline{\includegraphics[height=4.3cm]{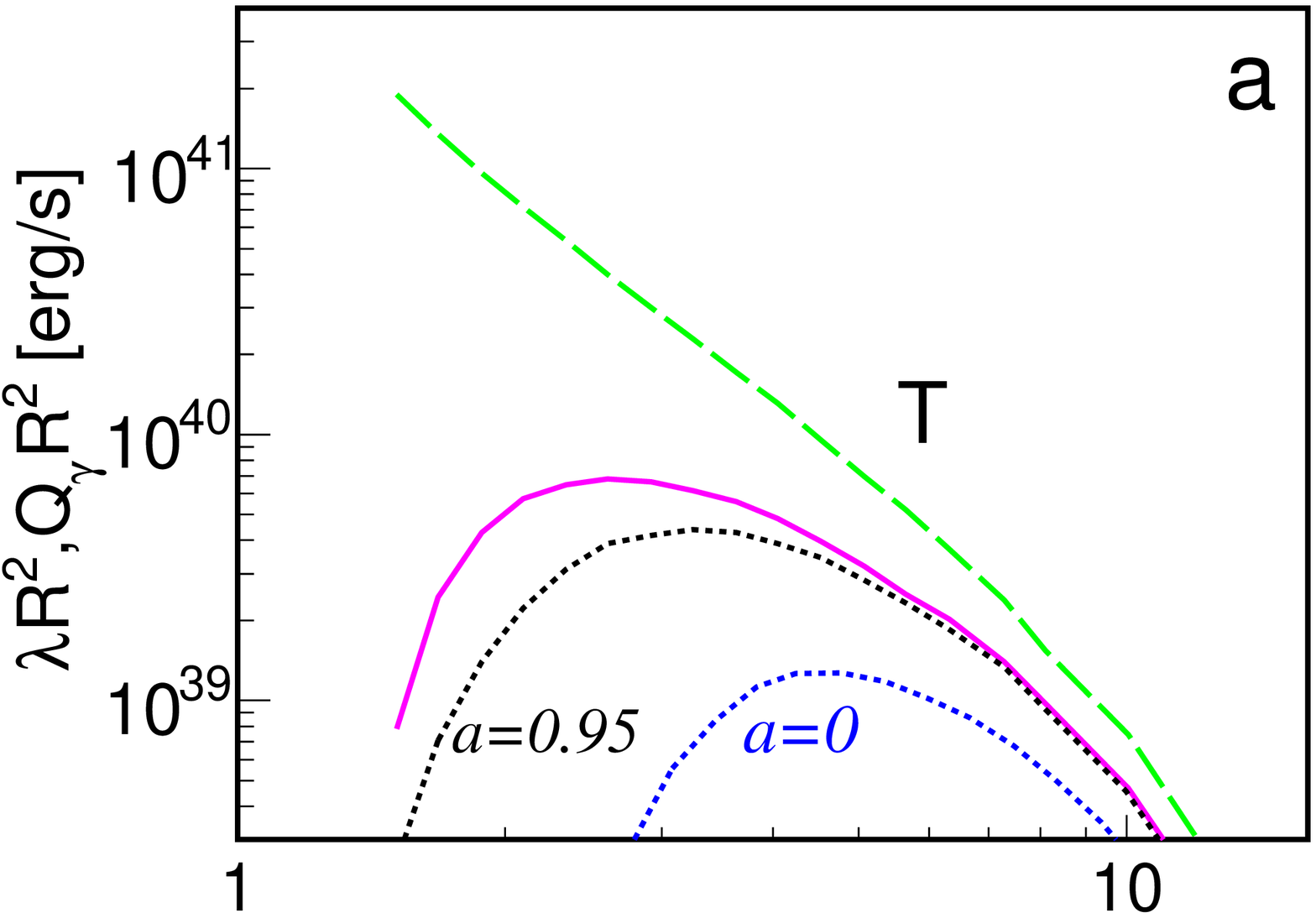} \includegraphics[height=4.3cm]{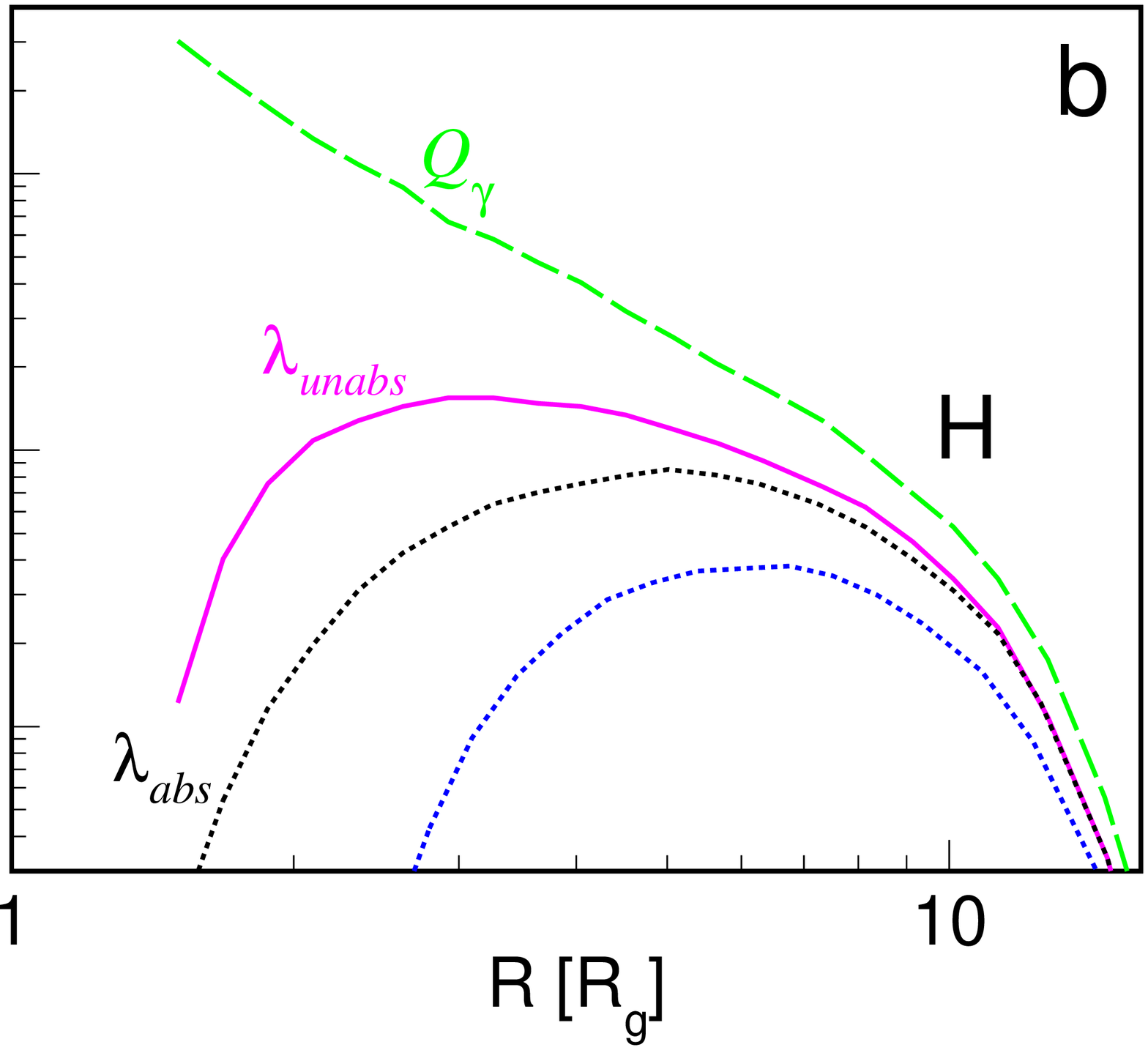}
\includegraphics[height=4.3cm]{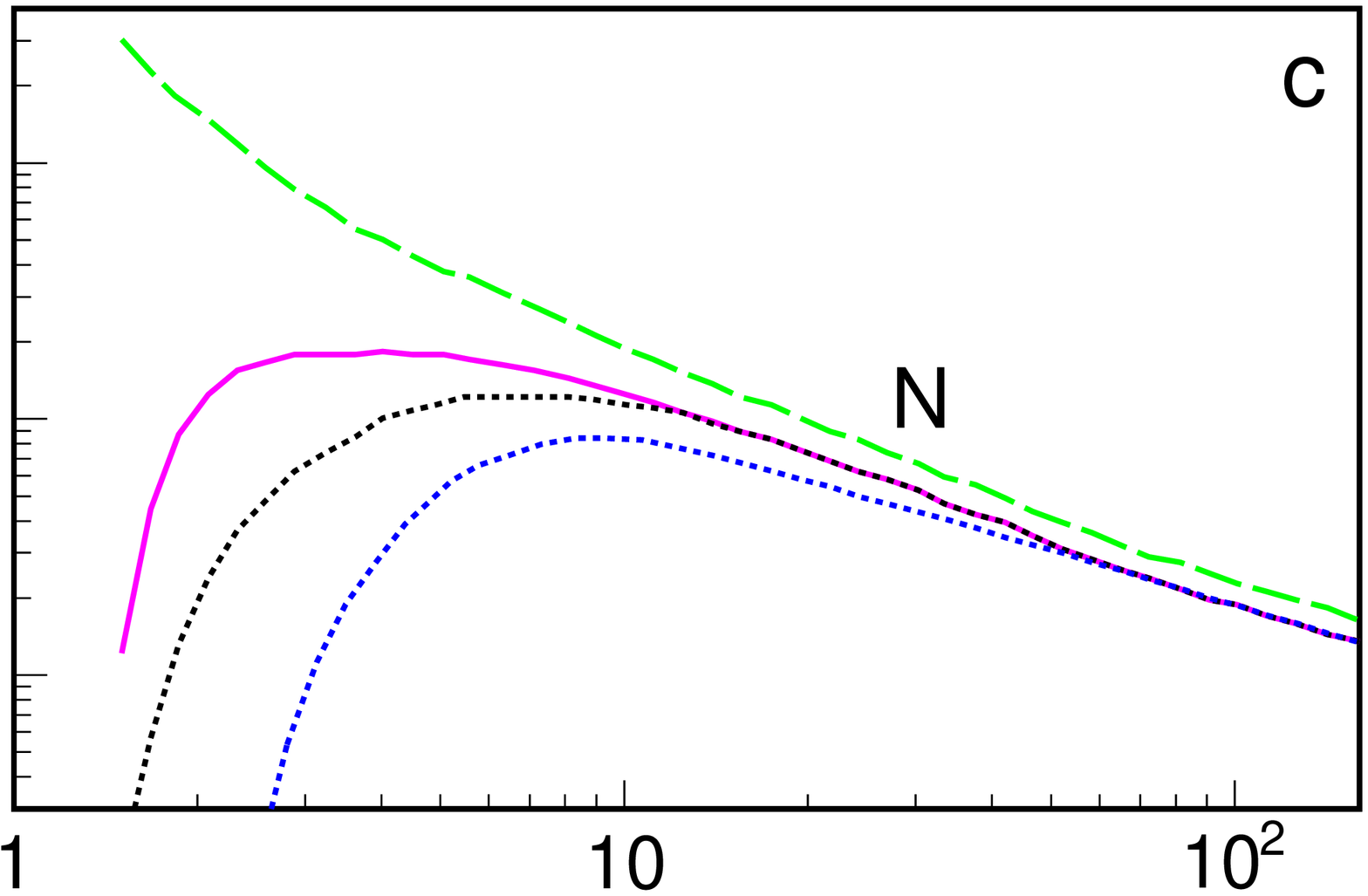}}
\caption{Dashed (green) lines show the radial profiles of the vertically-integrated $\gamma$-ray emissivities, $Q_{\gamma}$,  for models with $a=0.95$. Solid (magenta)  lines show the local contribution to  the luminosity at infinity from a unit area of the flow neglecting  the   $\gamma \gamma$ absorption, $\lambda_{\rm unabs}$, for $a=0.95$. The dotted lines show  the local contribution to  the luminosity at infinity from a unit area of the flow taking into account  the   $\gamma \gamma$ absorption, $\lambda_{\rm abs}$, for $a=0.95$ (upper, black) and $a=0$ and (lower, blue). (a) model T; (b) model H with $s=2.6$; (c) model N with $s=2.6$.  All models assume $\delta=10^{-3}$. 
}
\label{fig:7} 
\end{figure*}

\section{Hadronic $\gamma$-ray emission and relativistic transfer effects}

The hydrodynamical  solutions set the proton density, $n_{\rm p}$, and temperature, $T_{\rm p}$, as a function of radius. In principle, this should allow to determine the $\gamma$-ray emissivity, resulting from neutral pion production in proton-proton collisions and their subsequent decay into $\gamma$-ray photons, in the rest frame of the flow. However, details of this process are subject to an uncertainty related to the distribution of proton energies, which is unlikely to be thermal in  optically thin flows (see discussion in Section \ref{sec:protons}). Following M97 and OM03 we assume that the temperature from the global solution functions as a parameter specifying the average energy of  protons in the plasma which, however, does not have to have a thermal distribution. We consider several phenomenological models which must satisfy the obvious requirements that at each radius (1) the number density of protons equals $n_{\rm p}(r)$, determined by the global ADAF solution and (2) the average energy of protons equals the average energy 
\begin{equation} 
U_{\rm th}(\theta_{\rm p}) = \theta_{\rm p} m_{\rm p}c^2 (6+15 \theta_{\rm p})/(4+5 \theta_{\rm p}). 
\label{eq:uth}
\end{equation} 
of the Maxwellian proton gas with temperature, $T_{\rm p}(r)$, determined by the global ADAF solution,  where $\theta_{\rm p} = kT_{\rm p}/m_{\rm p}c^2$ and we use the simplified (cf.\ Gammie \& Popham 1998)  relativistic form of $U_{\rm th}(\theta_{\rm p})$.

We consider models involving various combinations of thermal 
\begin{equation} 
n_{\rm th}(\gamma) = n_{\rm th} \gamma^2 \beta \exp(-\gamma/\theta_{\rm p})/[\theta_{\rm p} K_2(1/\theta_{\rm p})] , 
\label{eq:thdist}
\end{equation}
and power-law 
\begin{equation} 
n_{\rm pl}(\gamma) = n_{\rm pl} (s-1) \gamma^{-s}, 
\label{eq:pldist}
\end{equation}
distributions of proton energies, where $n_{\rm th}$ and  $n_{\rm pl}$ are the local densities of these two populations. The thermal model (model T) assuming a purely Maxwellian distribution of protons and the nonthermal model (model N, same to nonthermal models of M97 and OM03), assuming that the total energy is stored in the power-law distribution of a small fraction of protons, allow us to estimate the {\it minimum} and {\it maximum} level of $\gamma$-ray luminosity, respectively, for a given set of ($M$, $\dot m$, $a$, $\alpha$, $\beta$, $\delta$). Mahadevan (1999) and OM03 considered the model involving the mixture of the thermal and power-law distributions, with the radius-independent parameter characterizing the fraction of energy that goes into the two distributions. Deviations of such a model from model N are trivial, with the $\gamma$-ray luminosity linearly proportional to the fraction of energy going to the power-law distribution. In this work we consider a different hybrid model (model H) with the radius dependent normalization between the power-law and the thermal distribution, which allows us to illustrate some additional effects. 

The detailed assumptions on the parameters of these models are as follows ($n_{\rm p}(r)$ and $T_{\rm p}(r)$ denote values given by the global ADAF solution):  

\noindent {\bf Model T} assumes a purely Maxwellian distribution of protons, equation (\ref{eq:thdist}), with $n_{\rm th}=n_{\rm p}(r)$ and $\theta_{\rm p}=kT_{\rm p}(r)/m_{\rm p}c^2$.

\noindent
{\bf Model N} assumes that a fraction $\psi$ of protons form the power-law distribution, equation (\ref{eq:pldist}), with the radius-independent index $s$ and $n_{\rm pl}= \psi(r) n_{\rm p}(r)$, and the remaining protons are cold, with the Lorentz factor $\gamma \simeq 1$ ($\gamma = 1$ is assumed in the computations). The radius-dependent fraction $\psi$ is determined by
\begin{equation}
{\psi m_{\rm p}c^2  \over s -2}  = U_{\rm th}\left[T_{\rm p}(r)\right].
\label{psi1}
\end{equation}

\noindent
{\bf Model H} assumes that an efficient nonthermal acceleration operates only within the central $\sim 15 R_{\rm g}$, where the average proton energies resulting from the ADAF solutions become relativistic. Specifically, we assume that at each radius at $r<15$ a fraction of protons form a thermal distribution at a subrelativistic temperature of $T = 4.3 \times 10^{11}$ K ($\theta_{\rm p} = 0.04$), and the remaining form a power-law distribution (equation \ref{eq:pldist}) with a constant (i.e.\ radius-independent) index $s$ and $n_{\rm pl}= \psi(r) n_{\rm p}(r)$. The relative normalization of these two distributions is determined by
\begin{equation}
{\psi m_{\rm p}c^2  \over s -2} + (1-\psi) {6.6 \over 4.2} m_{\rm p}c^2 =  U_{\rm th}\left[T_{\rm p}(r)\right],
\label{psi2}
\end{equation}
(where the factor $6.6 / 4.2$ results from equation (\ref{eq:uth}) with $\theta_{\rm p}=0.04$). At $r>15$, where $T_{\rm p}<4.3 \times 10^{11}$ K, there are no non-thermal protons in this model, which then results in a negligible pion production at such distances, similar as in model T. The chosen value of $T = 4.3 \times 10^{11}$ K gives a smooth transition between a purely thermal and a hybrid plasma at $r = 15$, however, radiative properties of model H are roughly independent of the specific value of the temperature of the subrelativistic thermal component. We remark also that $T=4.3 \times 10^{11}$ K is close to the limiting temperature above which the pion production prevents thermalization of protons (see Stepney 1983, Dermer 1986b)

The efficiency of pion production by protons with the power-law distribution increases with the decrease of the power-law index $s$. On the other hand, the fraction $\psi$ decreases with decreasing $s$, roughly as $\psi \propto (s-2)$. These two effects balance each other yielding the largest luminosity in 0.1--10 GeV range,  $L_{\rm 0.1-10 GeV}$, for $s \simeq 2.5-2.6$. For $2.3 < s < 2.8$, the dependence of $L_{\rm 0.1-10 GeV}$ on $s$ is weak; for $s = 2.1$,  $L_{\rm 0.1-10 GeV}$ is by a factor of $\sim 2$ smaller than for $s=2.6$. To estimate the maximum value of $L_{\rm 0.1-10 GeV}$ that can be produced in a flow with given parameters, in our computations for models N and H we use $s=2.6$. For all values of $a$, $\theta_{\rm p}>0.1$, and also $\psi>0.1$ in models H and N with $s=2.6$, within the innermost several $R_{\rm g}$.

In our solutions  of the flow structure we assume that protons are thermal and we use the thermal form of the gas pressure. Then, our models N and H with non-thermal proton distributions are not strictly self-consistent, as their pressure may deviate from the thermal prescription. However, this is a rather small effect, e.g.\ the pressure of the purely non-thermal distribution (model N) differs by 20--30 per cent from the pressure of a thermal gas with the same internal energy.  

For a given distribution of proton energies we determine the  $\gamma$-ray spectra in the flow rest frame, strictly following Dermer (1986a,1986b), in a manner  similar to M97 and OM03; however, we do not apply the following simplification underlying their nonthermal model. As argued in M97, the fraction of nonthermal protons should be small, $\psi \ll 1$, and, therefore, interactions of nonthermal protons with other nonthermal protons may be neglected; hence, only interaction of nonthermal protons with cold protons are taken into account in their computations. We remark that such an approach underestimates the $\gamma$-ray luminosity, e.g.\ by a factor of $\sim 2$ in model N with $a=0.95$ and $s=2.6$ (for which $\psi \simeq 0.4$ in the innermost region). In all our models we take into account interaction of protons with all other protons.

To compute the $\gamma$-ray luminosity and spectra received by distant observers we use a Monte Carlo method similar to that described in Nied\'zwiecki (2005). We generate $\gamma$-ray photons isotropically in the plasma frame, make a Lorentz transformation from the flow rest frame to the locally non rotating (LNR) frame and then we compute the transfer of $\gamma$-ray photons in curved space-time; see, e.g., Bardeen et al. (1972) for the definition of LNR frames and the equations of motion in the Kerr metric.

\begin{figure} 
\centerline{\includegraphics[height=3.2cm]{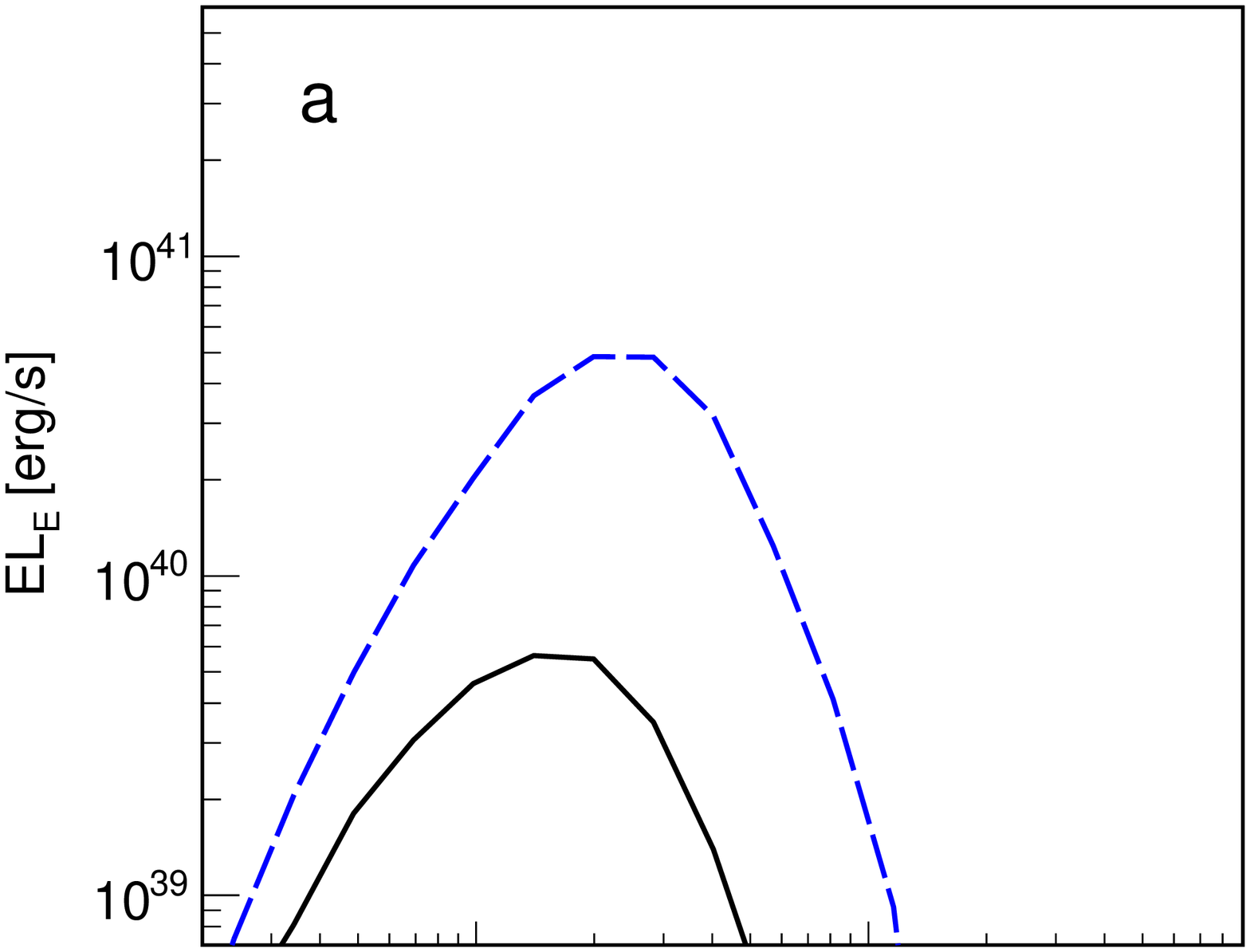} \includegraphics[height=3.2cm]{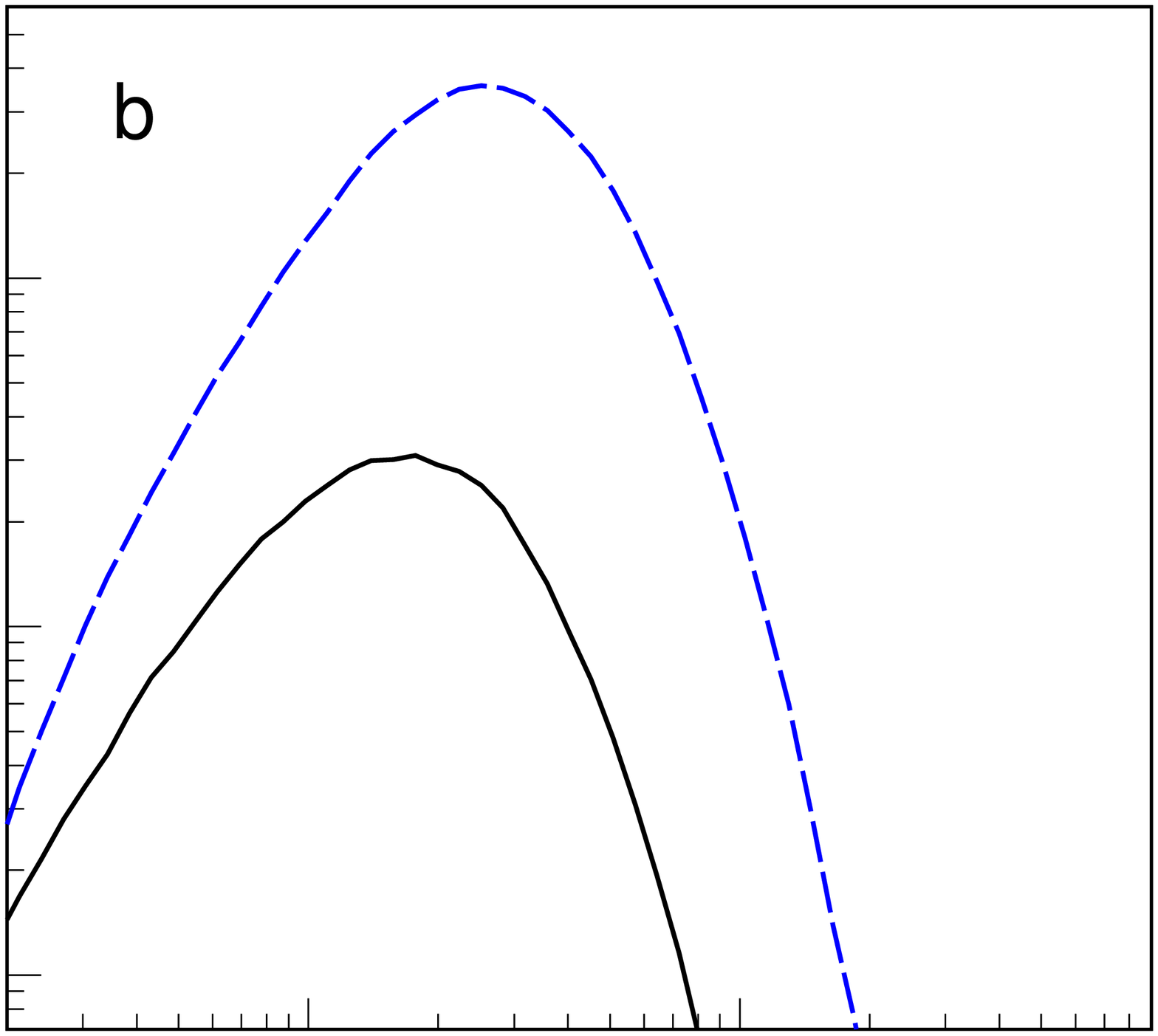}} 
\centerline{\includegraphics[height=3.5cm]{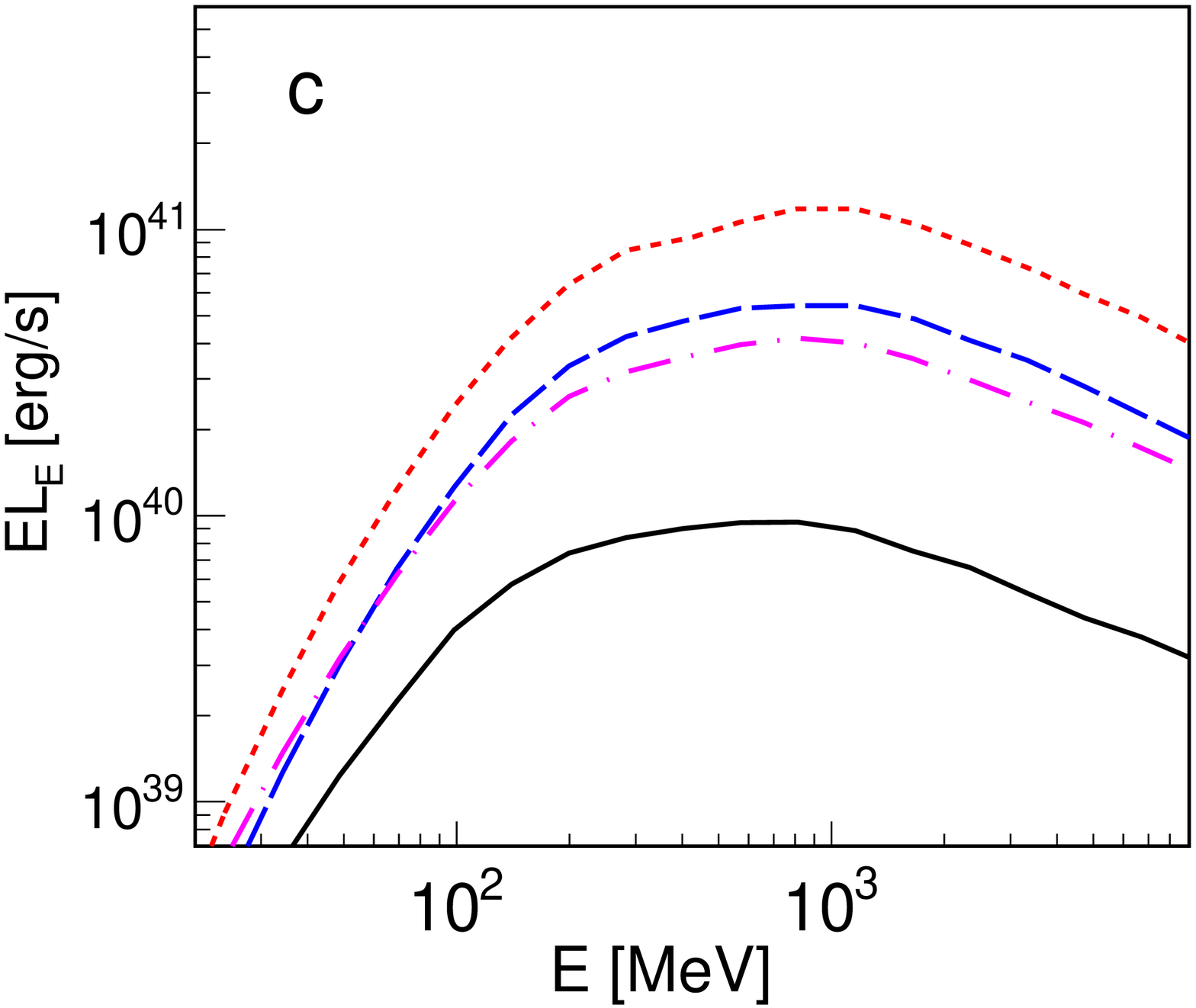} \includegraphics[height=3.5cm]{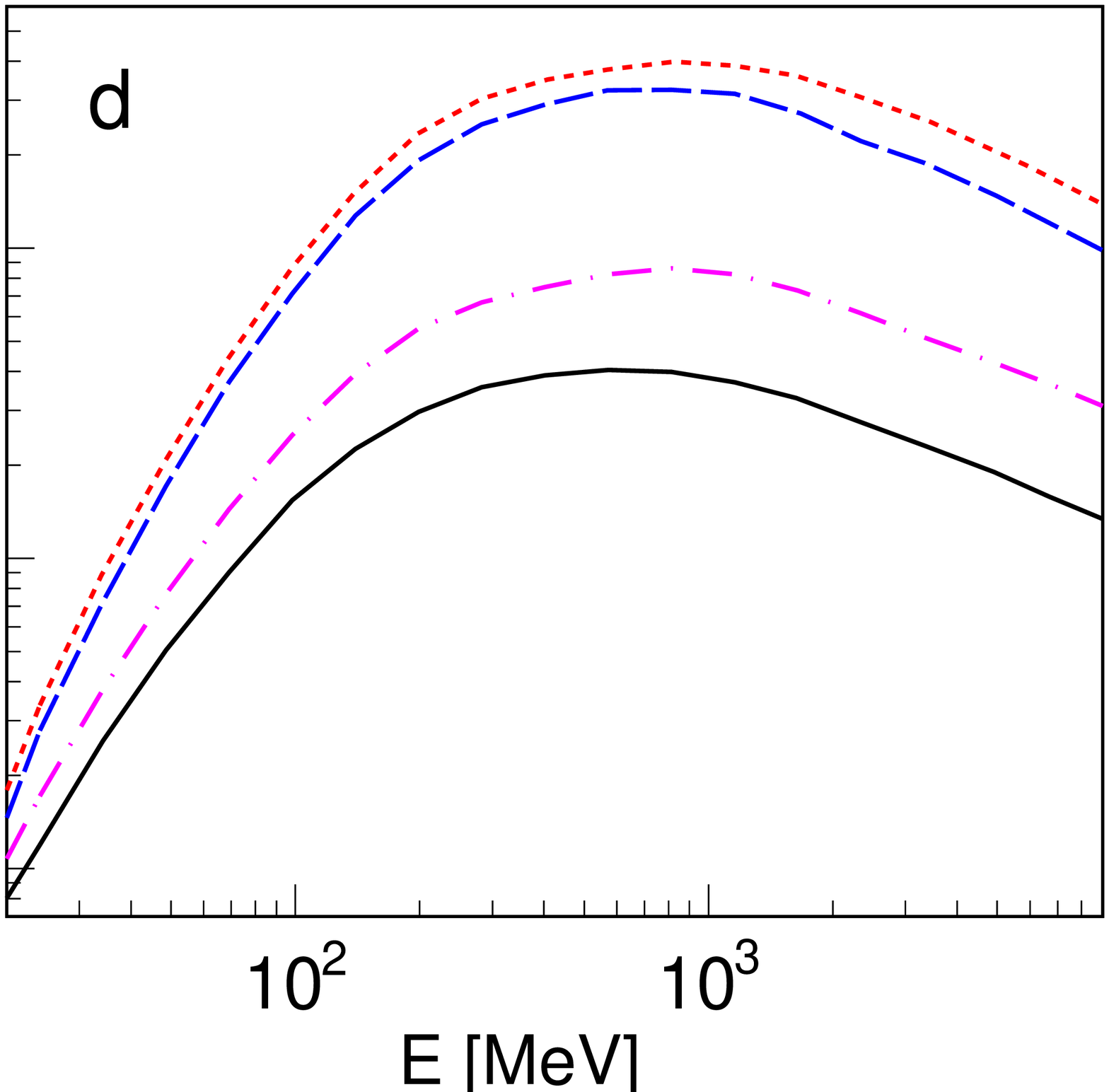}} 
\caption{Dashed (blue) and solid (black) lines show the rest frame and the observed $\gamma$-ray spectra, respectively,  for model T (ab) and model H with $s=2.6$ (cd). Dotted (red) and dot-dashed (magenta) lines in (cd) show the rest frame and the observed $\gamma$-ray spectra, respectively,  for model N with $s=2.6$. All spectra are for $\delta=10^{-3}$; panels (a) and (c) are for $a=0$, panels (b) and (d) are for  $a=0.95$. In this figure, the observed spectra neglect $\gamma \gamma$ absorption, so they are effected only by GR effects.
}
\label{fig:3} 
\end{figure}

The dashed lines in Fig.\ \ref{fig:7} show the radial profiles of the vertically-integrated $\gamma$-ray emissivity, $Q_{\gamma}$ ($Q_{\gamma}$ gives the energy emitted from the unit area per unit time)  for models T, H and N with $a=0.95$. The solid lines in Fig.\ \ref{fig:7} show  the radial profiles of the  vertically-integrated local luminosity (the energy per unit time reaching infinity from the unit area at a given $r$). The local luminosity profiles shown by the solid lines neglect the $\gamma \gamma$ absorption, so the difference between the dashed and solid lines is only due to the relativistic transfer effects.  Fig.\ \ref{fig:3} shows the corresponding $\gamma$-ray spectra and compares them with the spectra for $a=0$. At $r<10$ both models N and H are characterised by similar values of $\psi$ and produce similar amounts of $\gamma$-rays. In both models T and H the contribution from $r > 10$ is very weak; in model N the radial emissivity is much flatter despite $\psi$ being small, e.g.\ $\psi < 5 \times 10^{-3}$ at $r > 100$. Comparing models T and H we can see the effect of the local proton distribution function and by comparing models H and N we can see the effect of the radial emissivity.

For the thermal distribution of protons, the rest-frame photon spectra are symmetrical, in the logarithmic scale, around $\sim 70$ MeV but in $E F_E$ units  they peak around 200 MeV; the position of the maximum in the spectra observed by distant observes is slightly redshifted. Note that the difference of $\gamma$-ray luminosities, $L_{\gamma}$, between $a=0$ and 0.95 in our model T is much smaller than that derived by OM03, whose thermal models with $a=0$ and 0.95 give  $L_{\gamma}$ differing by approximately three orders of magnitude.  The difference is due to different values of $\beta_{\rm B}$ assumed here and by OM03, which result in different $\theta_{\rm p}$, as discussed in Section \ref{sec:flow}.
The dependence of $L_{\gamma}$ on $\theta_{\rm p}$ changes around $\theta_{\rm p} \approx 0.1$  (see, e.g.,  fig.\ 3 in Dermer 1986b). At lower temperatures, $L_{\gamma}$ is extremely sensitive to $\theta_{\rm p}$, with the increase of $\theta_{\rm p}$ by a factor of 2 yielding the increase of $L_{\gamma}$ by over two orders of magnitude. At $\theta_{\rm p}>0.1$, the dependence is more modest, e.g.\ the increase  of $\theta_{\rm p}$ from 0.2 to 0.4 results in the increase of $L_{\gamma}$ by only a factor of $\sim 2$.
For $\beta_{\rm B}=0.9$ assumed in this work, $\theta_{\rm p} > 0.1$ at small $r$ for all values of $a$, making the $\gamma$-ray luminosity much less dependent on the black hole spin. For  $\beta_{\rm B}=0.5$,
assumed by OM03,     the proton temperature is small, with the maximum value of $\theta_{\rm p} \approx 0.03$ for $a=0$, which leads to the strong dependence of   $L_{\gamma}$ on $a$.

For both model H and N, the spectrum at $E > 1$ GeV has the same slope as the power-law distribution of proton energies. For model H with $s=2.6$,  $L_{\gamma}$ is by a factor of 3 larger than in model T. Rather small difference between $L_{\gamma}$ in our thermal and nonthermal models is again due to our assumption of a weak magnetic field. At smaller $\beta_{\rm B}$, resulting in smaller $\theta_{\rm p}$, the presence of even a small fraction of non-thermal electrons leads to the increase of $L_\gamma$ 
by orders of magnitude, as can be seen by comparing the  emissivities of our models N and T at $r > 10$ (see also M97).

\begin{figure} 
\centerline{
\includegraphics[height=3.5cm]{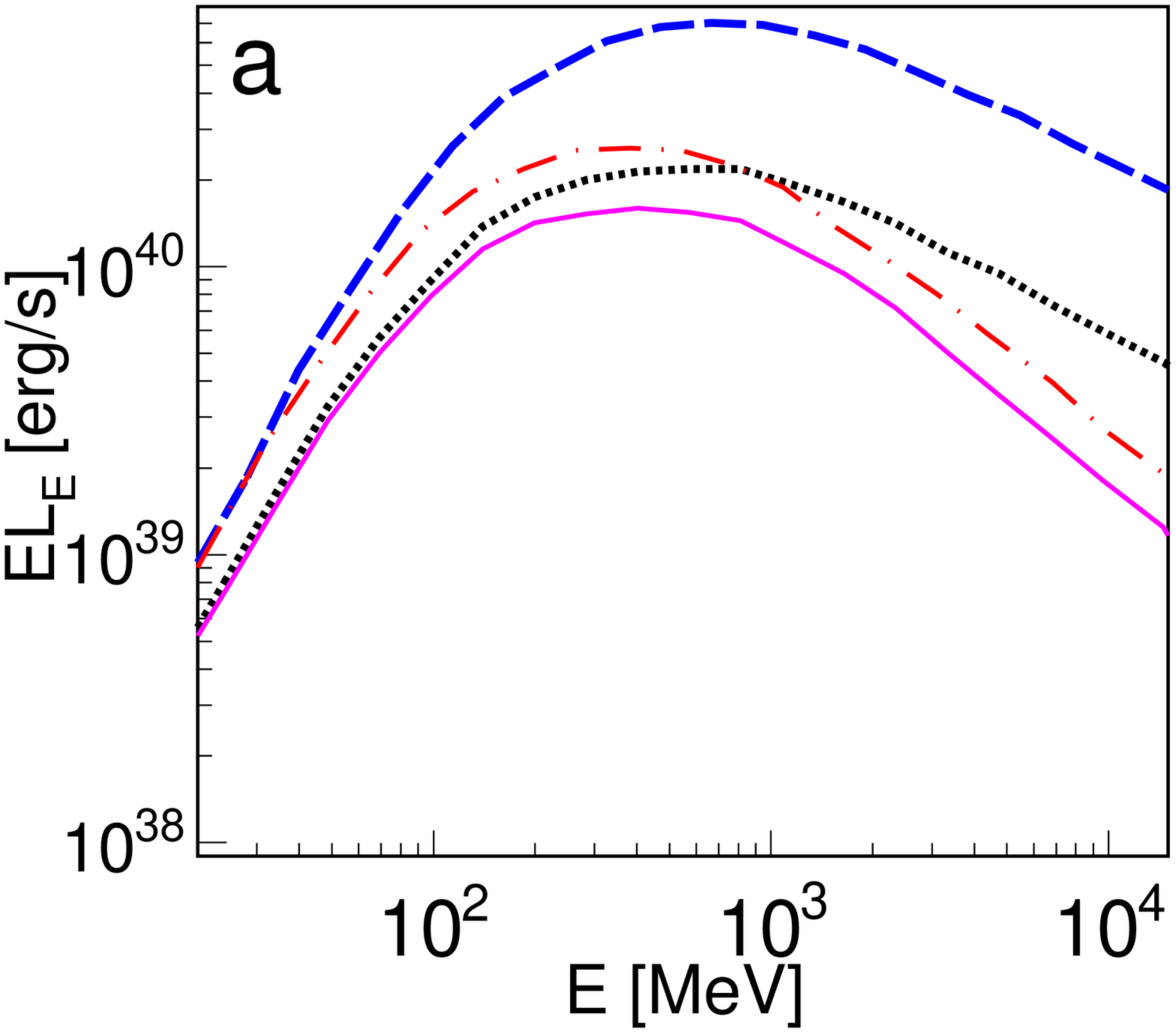} \includegraphics[height=3.5cm]{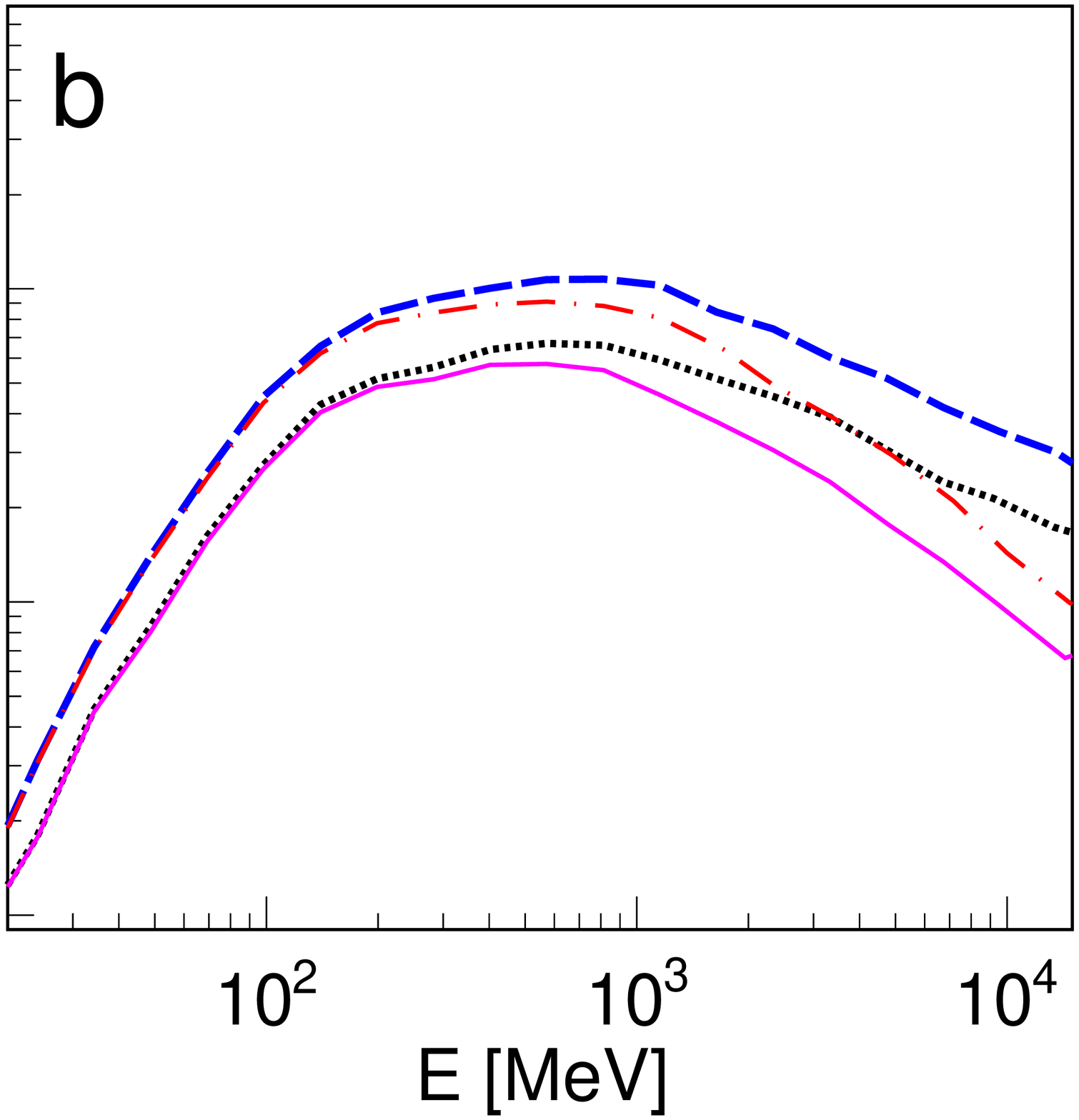}} 
\caption{Observation-angle dependent $\gamma$-ray spectra taking into account and neglecting the $\gamma \gamma$ absorption for model H with $s=2.6$; $a=0.998$ (a) and $a=0$ (b). The dot-dashed (red) and dashed (blue) lines show the spectra observed at $\theta_{\rm obs}=70^{\rm o}$, with and without absorption, and the solid (magenta)  and dotted (black) lines show the spectra observed at $\theta_{\rm obs}=30^{\rm o}$, with and without absorption, respectively.
}
\label{fig:8} 
\end{figure}

For models T and H the bulk of the $\gamma$-ray emission comes from $r < 10$ (Fig.\ \ref{fig:7}ab) and the GR transfer effects reduce
the detected $\gamma$-ray flux by approximately an order of magnitude.
In model N the magnitude of the GR effects on the total flux is reduced due to strong contribution from $r>10$ 
(which is weakly affected by GR). Also in model N, the contribution from $r>10$, which for $a=0$ 
approximately equals the contribution from $r<10$, reduces the difference between the 
$\gamma$-ray fluxes observed for $a=0$ and $a=0.95$ to only a factor of $\sim 2$, see Fig.\ \ref{fig:3}(cd).

\begin{figure*}
\centerline{\includegraphics[width=9.4cm]{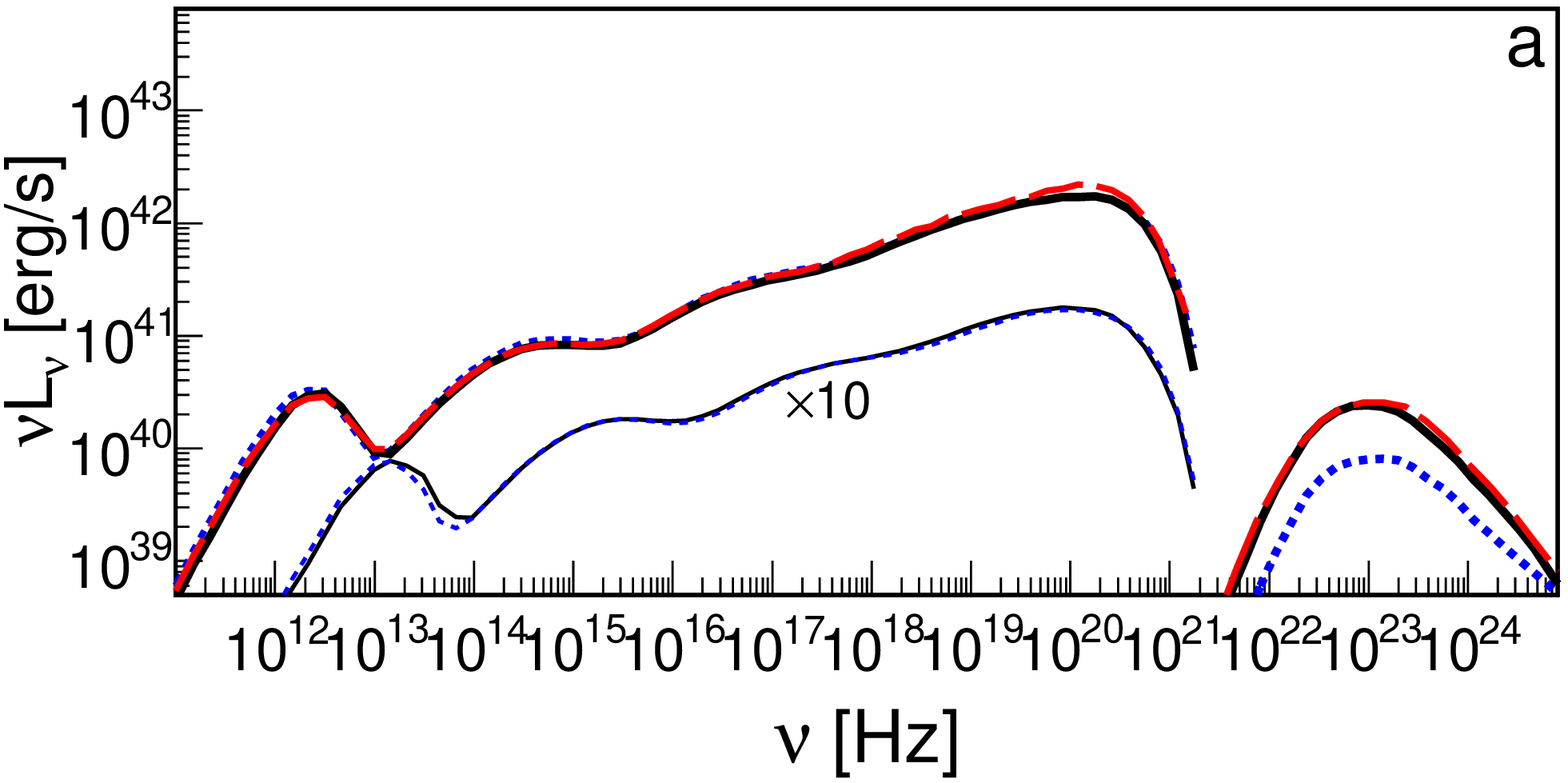}\includegraphics[width=8.1cm]{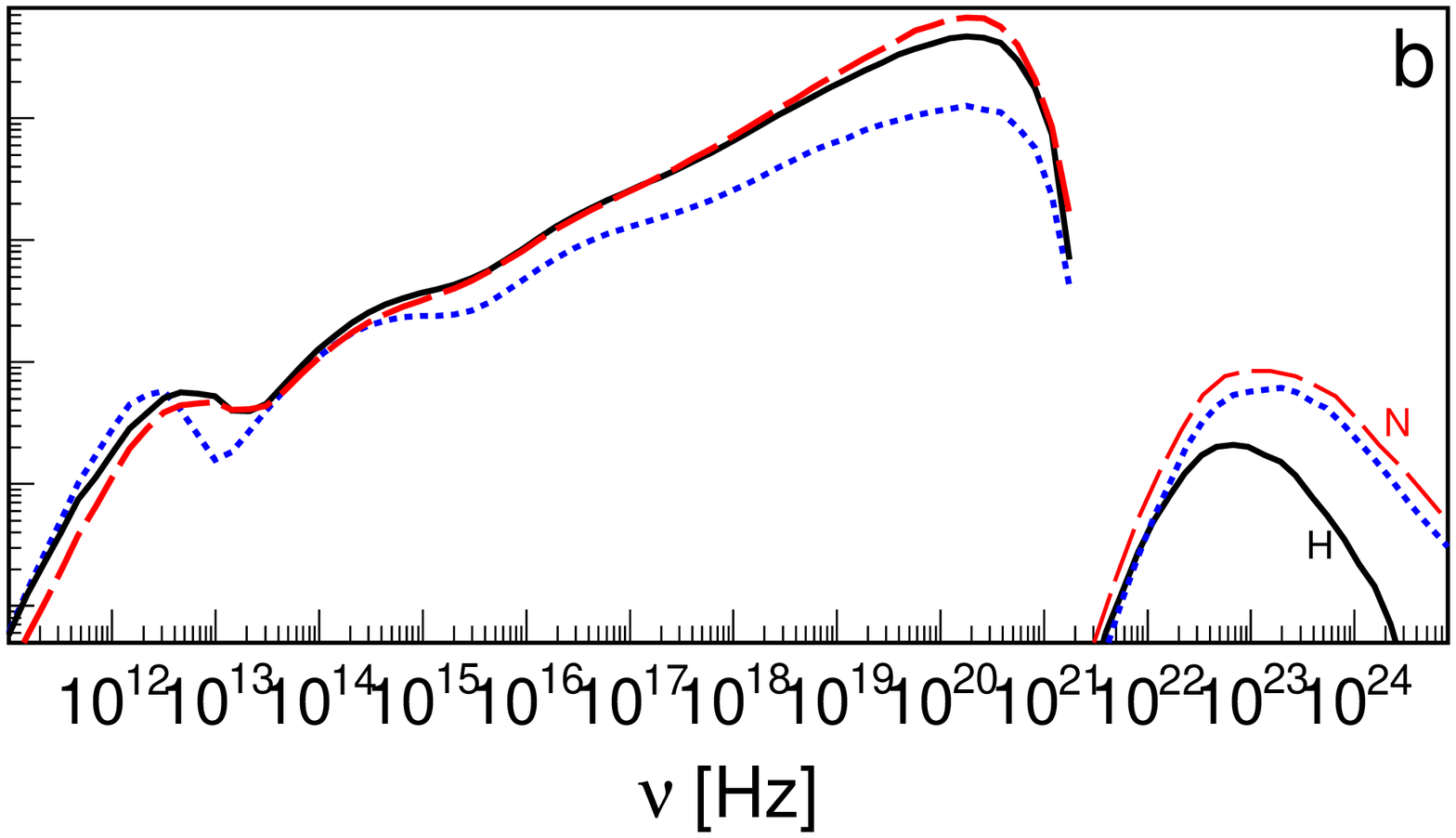}}
\caption{Angle-averaged spectra received by a distant observer; the synchrotron and Comptonized (radio to X-rays) and hadronic ($\gamma$-rays) components are shown separately. In both panels $a=0.998$ (dashed, red), 0.95 (solid, black) and 0 (dotted, blue) (a) Models with $\delta=10^{-3}$;  the $\gamma$-ray spectra correspond to model H with $s=2.6$. (b) Models with $\delta=0.5$; the $\gamma$-ray spectra correspond to model N with $s=2.6$ for $a=0$ and 0.998 and to model H with $s=2.6$ for  $a=0.95$. The lower pair of lines in panel (a), rescaled by a factor of 10, show the spectra (leptonic component) for $M =2 \times 10^6 M_{\odot}$.}
\label{fig:6} 
\end{figure*}

The viewing-angle dependent spectra for model H, which would be observed (if unabsorbed) by distant observers, are shown by the dashed and dotted lines in Fig.\ \ref{fig:8}. The flows considered in this work are quasi-spherical and optically thin and hence their appearance depends on the viewing angle primarily due to the relativistic transfer effects. Most importantly, trajectories of photons emitted close to a rapidly rotating black hole are bent toward its equatorial plane. Therefore, the $\gamma$-ray radiation has a significant intrinsic anisotropy in models with large $a$, with edge-on directions corresponding to larger $\gamma$-ray fluxes.




\section{Comptonization and $\gamma \gamma$ absorption}

The absorption of $\gamma$-rays in the radiation field of the flow has been calculated in a fully GR model by Li et al.\ (2009). Here we use a similar approach with the major difference involving the computation of target photon density. Li et al.\ (2009) considered the propagation of photons with energies of 10 TeV, which are absorbed mostly in interactions with infra-red photons. Those low energy photons are produced primarily by the synchrotron emission which can be simply modelled using its local emissivity. In turn, photons with energies in 0.1--10 GeV range, considered in this work, are mostly absorbed by the UV and soft X-ray photons, which are produced by Comptonization. Then, an exact  computation of the angular-, energy- and location-dependent distribution of the target photon field requires the precise modelling of the Comptonization taking into account its global nature. In our model we apply the Monte Carlo (MC) method, described in detail in N12, with seed photons for Comptonization from synchrotron and bremsstrahlung emission.

\begin{figure} 
\centerline{
\includegraphics[width=6cm]{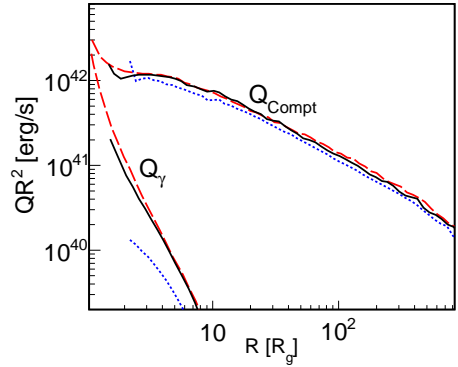}}
\caption{Radial profiles of the $\gamma$-ray emissivity, $Q_{\gamma}$ (for model T), and the Comptonization rate $Q_{\rm Compt}$, for  $a=0.998$ (dashed, red), 0.95 (solid, black) and 0 (dotted, blue) in models with $\delta=10^{-3}$. $Q$ denotes the vertically integrated rates.
}
\label{fig:4} 
\end{figure}

We find self-consistent electron temperature distributions using the procedure described in N12; we iterate between the solutions of the electron energy equation (analogous to equations \ref{eq:ions}--\ref{eq:qdiss}; note that here we include the direct electron heating, while N12 assumes $\delta=0$) and the GR MC Comptonization simulations until we find mutually consistent solutions. In Fig.\ \ref{fig:6} we show the resulting spectra. Fig.\ \ref{fig:4} shows the radial profiles of the radiative cooling of electrons (strongly dominated by Comptonization), $Q_{\rm Compt}$, for $\delta=10^{-3}$ and compares them with the $\gamma$-ray emissivity, $Q_{\gamma}$, for   model T. Note that $Q_{\gamma}$ is much steeper than $Q_{\rm Compt}$  so the GR effects are more important for the $\gamma$-ray than for the X-ray emission.

As we can see in Fig.\ \ref{fig:4} and also in the corresponding spectra in Fig.\ \ref{fig:6}a, for $\delta = 10^{-3}$ the black hole spin negligibly affects the Comptonized radiation; this property results from a large magnitude of the compression work, which is roughly independent of $a$ and dominates the heating of electrons for small values of $\delta$ (cf.\ N12).

For $\delta \ge 0.1$ the direct heating contributes significantly to the heating of electrons and for $\delta=0.5$ it dominates over other heating processes at $r<100$ for all values of $a$. Then, the dependence of  $Q_{\rm vis}$ on $a$ results in a noticeable dependence of the Comptonized radiation on $a$ for $\delta \ge 0.1$ (see also Xie \& Yuan 2012 for a recent study of the dependence of X-ray luminosity on $\delta$). The radiative efficiency increases from $\eta = 0.004$ for all values of $a$ at $\delta=10^{-3}$ to $\eta=0.02$ for $a=0$, $\eta=0.08$ for $a=0.95$  and $\eta=0.1$ for $a=0.998$ at $\delta=0.5$. Despite considering only one value of accretion rate, our solutions span a range of bolometric luminosities, from  $L \approx 4 \times 10^{-4} L_{\rm Edd}$ (for $\delta=10^{-3}$) to $L \approx 10^{-2} L_{\rm Edd}$ (for $\delta=0.5$ and $a = 0.998$). The corresponding  X-ray spectral slopes harden from $\Gamma_{\rm X} \simeq 1.7$ to  $\Gamma_{\rm X} \simeq 1.5$ with the increase of $L$. Note that these values correspond to the range of parameters close to the turning point in the $L$--$\Gamma$ correlation observed in AGNs (e.g.\ Gu \& Cao 2009). Then, we likely consider here the range of the largest luminosities of the flows in which synchrotron emission is the dominant source of seed photons for Comptonization (see discussion and references in N12).

Having found the self-consistent solutions, described above, we apply our MC model to tabulate the distribution of all photons propagating in the central region (up to $r_{\rm out}=1000$), ${\rm d}n_{\rm ph}(R,\theta,E_{\rm LN},\Omega_{\rm LN})/{\rm d}E_{\rm LN}{\rm d}\Omega_{\rm LN}$ (in photons cm$^{-3}$ eV$^{-1}$ sr$^{-1}$), where $R$ and $\theta$ are the Boyer-Lindquist coordinates, $E_{\rm LN}$ is the photon energy in the LNR frame and ${\rm d}\Omega_{\rm LN}$ is the solid angle element in the LNR frame.

To compute the optical depth to pair creation, $\tau_{\gamma\gamma}$, we closely follow the method for determining an optical depth to Compton scattering in the Kerr metric, see Nied\'zwiecki (2005) and  Nied\'zwiecki \& Zdziarski (2006), however, here we calculate the probability of pair creation in the LNR frame whereas for the Compton effect an additional boost to the flow rest frame is applied. While Compton scattering is most conveniently described in the plasma rest frame, pair production can be simply modelled in the LNR frame and, thus, the transformation to the flow rest frame is not necessary here. We solve equations of the photon motion in the Kerr metric and we determine the increase of the optical depth along the photon trajectory from
\begin{equation}
{\rm d}\tau_{\gamma\gamma} = \int \int \int (1 - \cos \theta_{\rm LN}) \sigma_{\gamma\gamma}
{ {\rm d}n_{\rm ph} \over {\rm d}E_{\rm LN}{\rm d}\Omega_{\rm LN} } {\rm d}E_{\rm LN}{\rm d}\Omega_{\rm LN}{\rm d}l_{\rm LN},
\label{eq:sgg}
\end{equation}
where ${\rm d}l_{\rm LN}$ is the length element in the LNR frame, $\sigma_{\gamma\gamma}(E_{\rm LN},E_{\rm \gamma LN},\theta_{\rm LN})$ is the pair production cross section (e.g., Gould \& Schreder 1967), $E_{\rm \gamma LN}$ is the energy of the $\gamma$-ray photon in the LNR frame and $\theta_{\rm LN}$ is the angle between the interacting photons in the LNR frame.

GR affects the $\gamma \gamma$ opacity through (1) bending the trajectories of both the $\gamma$-ray photon and target photons and (2) changing energies of both the $\gamma$-ray photon and target photons. As an example, the neglect of the gravitational shift of the $\gamma$-ray photon energy, by using $\sigma_{\gamma\gamma}(E_{\rm LN},E_{\gamma},\theta_{\rm LN})$ (where $E_{\gamma}$ is the energy at infinity) instead of $\sigma_{\gamma\gamma}(E_{\rm LN},E_{\rm \gamma LN},\theta_{\rm LN})$  in equation (\ref{eq:sgg}), underestimates $\tau_{\gamma\gamma}$ by a factor of $\approx 2$--3 for photons emitted from the innermost region.

\begin{figure} 
\centerline{
\includegraphics[width=6cm]{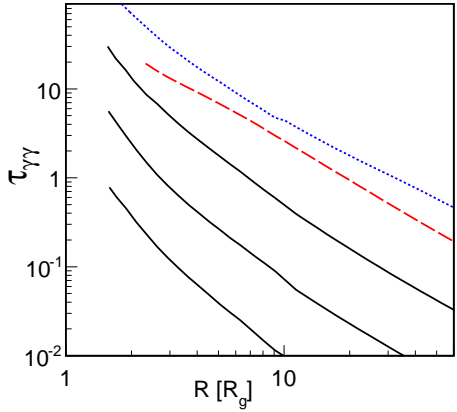}}
\caption{The optical depth to pair creation for radially outgoing $\gamma$-ray photons with $E_{\gamma}=100$ MeV, 1 GeV and 10 GeV from bottom to top, as a function of the radial distance of their point of emission for $a=0.95$ and $\delta=10^{-3}$ are shown by the solid (black) lines. The dotted (blue) and dashed (red) lines are for $a=0.95$ and 0 in models $\delta=0.5$; in these models  $\tau_{\gamma\gamma}$ is shown  only for $E = 10$ GeV for clarity.
}
\label{fig:5} 
\end{figure}

In Fig.\ \ref{fig:5} we show values of the total optical depth, $\tau_{\gamma\gamma}(r)$, integrated along the outward radial direction in the equatorial plane  from the emission point at the radial coordinate $r$ to the outer boundary at $r_{\rm out}$. As we can see, the $\gamma \gamma$ opacity is a strong function of both the $\gamma$-ray energy and the location in the flow. The dotted lines in Fig.\ \ref{fig:7} show how the $\gamma \gamma$ absorption attenuates $\gamma$-rays observed from a given $r$.

\begin{figure*} 
\centerline{\includegraphics[height=6.3cm]{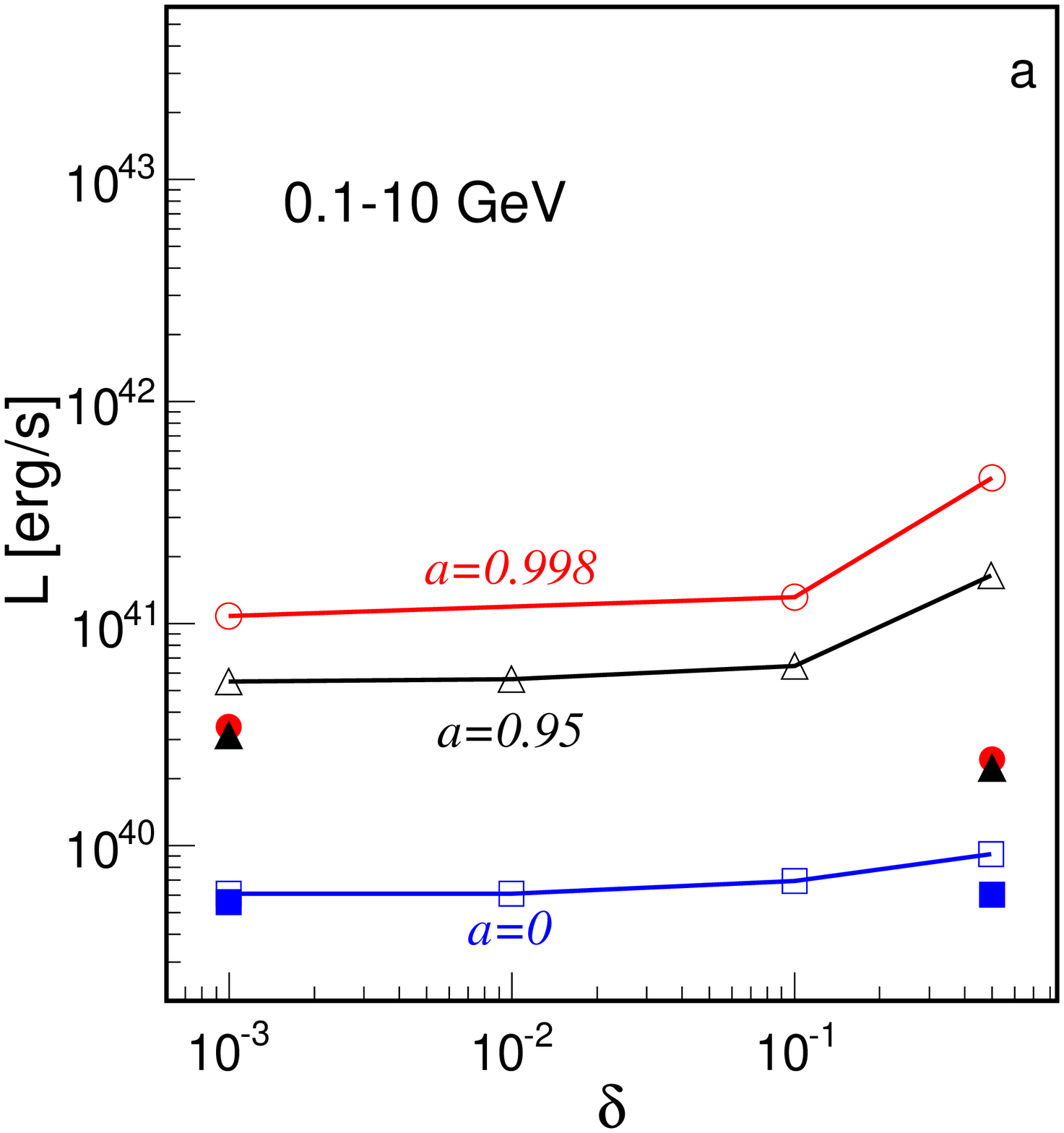}\includegraphics[height=6.3cm]{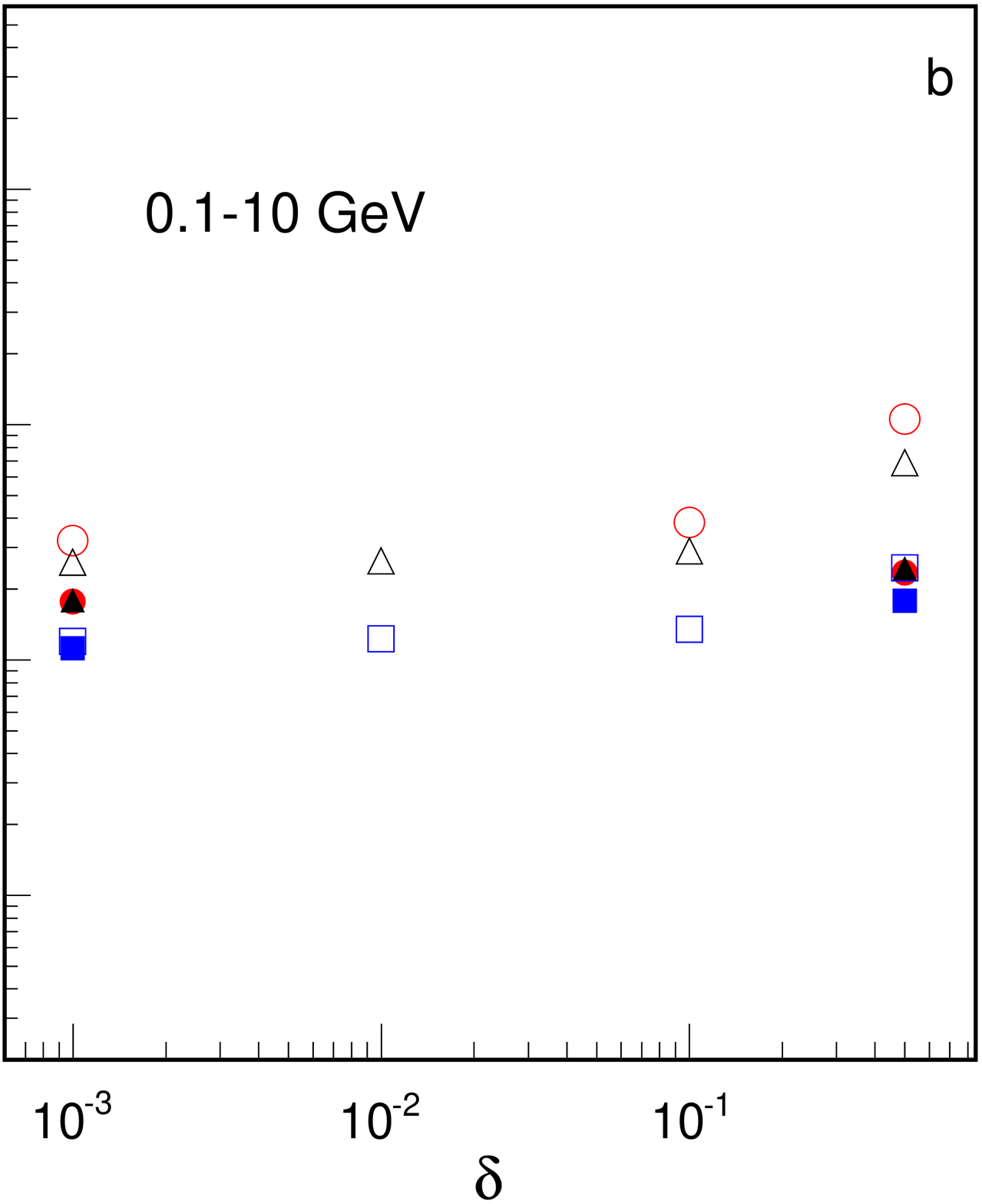}
\includegraphics[height=6.3cm]{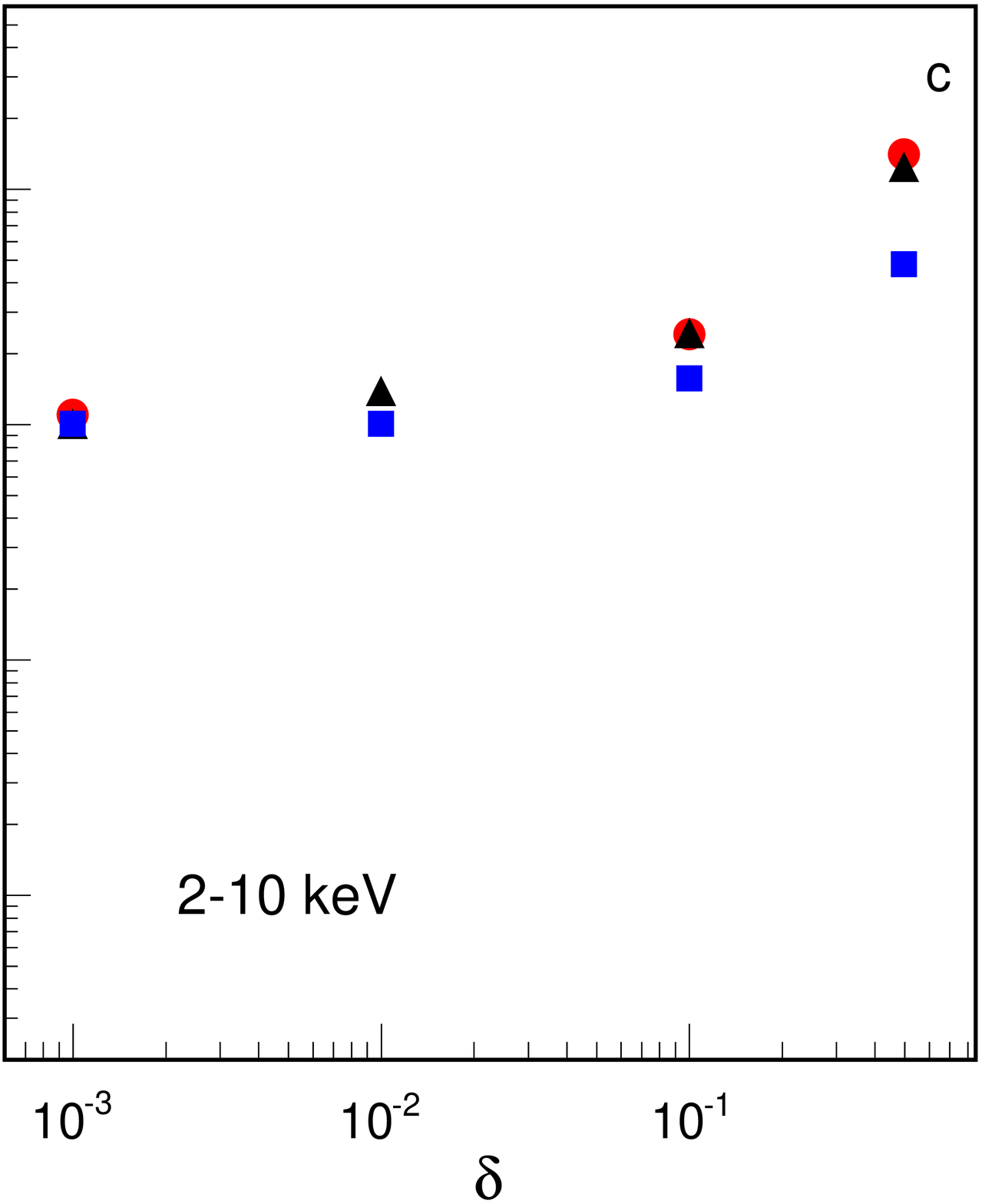}}
\caption{Observed, angle-averaged $\gamma$-ray luminosity in 0.1--10 GeV range in model T (a) and model N with $s=2.6$ (b) and X-ray luminosity in the 2--10 keV range (c) as a function of $\delta$ for $a=0.998$ (circles), $a=0.95$ (triangles) and $a=0$ (squares). The filled symbols in (ab) show $L_{\rm 0.1-10 GeV}$ after the $\gamma \gamma$ absorption, the open symbols show the luminosity $L_{\rm unabs}$ neglecting the absorption.  
}
\label{fig:9} 
\end{figure*}

It is apparent that around $\dot m \sim 0.1$ flows undergo transition from being fully transparent to mostly opaque to $\gamma$-rays. In our models with the Eddington ratio $L/L_{\rm Edd}= 4 \times 10^{-4}$, the flow is fully transparent to photons with energies $\la 100$ MeV; at higher energies the absorption leads to moderate attenuation,  with the increase of the photon index at $E > 1$ GeV by $\Delta \Gamma \simeq 0.2$, see Fig.\ \ref{fig:8}. The size of the $\gamma$-ray photosphere (the surface of $\tau_{\gamma\gamma}=1$) increases with increasing $L$ and for $L \simeq 10^{-2} L_{\rm Edd}$ the GeV photons cannot escape from $r<10$. At such $L$, our model H gives spectra with a clear cut-off around 1 GeV (see the solid line in Fig.\ \ref{fig:6}b)  which could be measured by {\it Fermi}. In other cases absorption leads to a smooth softening of the spectra.  

In terms of the 2-10 keV luminosity, $L_{\rm 2-10 keV}$, flows with $L_{\rm 2-10 keV} < 10^{-5} L_{\rm Edd}$ should be fully transparent to MeV/GeV photons. Flows with $L_{\rm 2-10 keV} > 10^{-3} L_{\rm Edd}$ can emit significant amounts of unabsorbed $\gamma$-rays only if their $\gamma$-ray emissivities are strong at large $r$. E.g.\ in our model N, the luminosity of the flow at $r>50$, which region would be outside the photosphere of 1 GeV photons even at much larger $L_{\rm 2-10 keV} \sim 10^{-2} L_{\rm Edd}$,  is $L_{\rm 0.1-10 GeV} \simeq 10^{40}$ erg/s. Then, the $\gamma$-ray luminosity exceeding $10^{41}$ erg/s can be expected at $\dot m > 0.3$ if the $\gamma$-ray emitting flow extends out to several tens of $R_{\rm g}$, which property is, however, unclear as objects with high luminosities often show signs of a cold disc extending to rather small radii (so the transition between the hot and cold flow may occur within the $\gamma$-ray photosphere). Note that for such a scenario, with $\gamma$-ray emission from a hot flow at large $L$, we expect a small luminosity ratio of $L_{\rm 0.1-10 GeV}/L_{\rm 2-10 keV} \sim 10^{-3}$ regardless of the value of $a$.


\section{X-ray  vs $\gamma$-ray luminosity}
\label{sec:6}

In Fig.\ \ref{fig:9} we summarize our results regarding the relation between the X-ray and $\gamma$-ray luminosities. The range of expected $L_{\rm 0.1-10 GeV}$ is constrained from below by values indicated in Fig.\ \ref{fig:9}a for model T, and from above by values in Fig.\ \ref{fig:9}b for model N. As we can see, the models give the luminosity ratios $L_{\rm 0.1-10 GeV}/L_{\rm 2-10 keV}$ between    $\sim 0.002$ and 0.2. 
 
In model T, $L_{\rm 0.1-10GeV}$ for $a=0$ and $a=0.998$ differ by a factor of several; the unabsorbed luminosities differ by over an order of magnitude but for $L$ close to $10^{-2}L_{\rm Edd}$ the $\gamma \gamma$ absorption reduces the difference to a factor of $\sim 4$. In model N, $L_{\rm 0.1-10GeV}$ for $a=0$ and $a=0.998$ differ by only  a factor of $\sim 2$; as noted before, the difference is reduced here due to contribution from large $r$. Model N for $a=0$ gives similar $L_{\rm 0.1-10GeV}$ as model T with large $a$; larger density and average energy for large $a$ is approximately compensated by a larger fraction of protons above the pion production threshold for model N. Note, however, that - despite similar luminosities - the spectra for these two regimes differ significantly, see Fig.\ \ref{fig:3}.

In model H,  $L_{\rm 0.1-10GeV}$ has a similar dependence on $a$ as in model T, with a large difference between small and high values of $a$. We conclude that it is the radial distribution of $\gamma$-ray emissivity, rather than the local proton distribution function, which reduces the dependence on $a$. This is even more hindering for attempts of assessing the spin value basing on the $\gamma$-ray luminosity level. The proton energy distribution function is reflected in the produced spectral shape and, therefore, it could be constrained observationally and taken into account in this kind of analysis. On the other hand, one cannot expect to derive information on the radial emissivity profile from observations, so an investigation of black hole spin values using the $\gamma$-ray luminosity should be subject to significant uncertainty. 

In models with a dominant contribution from the central $\sim 10 R_{\rm g}$, intrinsic $\gamma$-ray luminosities of flows around submaximally ($a=0.95$) and maximally  ($a=0.998$) rotating black holes differ by a factor of $\sim 2$. For the latter, the flow extends to smaller radii  and hence a larger proton temperature and density are achieved, see Fig.\ \ref{fig:1}. However, a strong contribution from $r > 10$ as well as $\gamma \gamma$ absorption make the difference between the observed luminosities insignificant.

Regarding the dependence on $\delta$, we notice a somewhat surprising property of models with $\delta=0.5$, which predict a larger $\gamma$-ray luminosity than models with smaller $\delta$ (the effect is more pronounced for larger values of $a$). The physical reason is that for $\delta=0.5$, the slight decrease of proton temperature is outweighed by the increase of density (through the decrease of both the radial velocity and the scale height with decreasing temperature), cf.\ Manmoto (2000), which leads to the increase of $L_\gamma \propto n_{\rm p}^2 T_{\rm p}$. At smaller values of $\delta$, the intrinsic $L_\gamma$ depends negligibly on $\delta$, in agreement with the results of OM03 for $a=0$ and $\delta \le 0.3$ in their nonthermal model.

The scaling of density with $\dot m$ and $M$ in our global GR solutions only weakly differs from  that of self-similar ADAF model,
i.e.\ $n \propto \dot m/M$ (e.g.\ Mahadevan 1997). Then, the intrinsic $\gamma$-ray luminosity can be estimated as $L_{\rm 0.1-10 GeV} \simeq L_{\rm unabs} (\dot m/0.1)^{2}(M/2 \times 10^8 M_\odot)^{-1}$ erg/s, where $L_{\rm unabs}$ is the unabsorbed luminosity shown by the open symbols in Fig.\ \ref{fig:9}(ab). This gives also the observed luminosity at $\dot m < 0.1$, when the absorption effects are unimportant. Obviously, the above scaling with $\dot m$ is not relevant for $\dot m \ga 0.1$, for which the increase of $\dot m$ results in the decrease of the observed $L_{\rm 0.1-10 GeV}$ due to $\gamma \gamma$ absorption. On the other hand,  the linear scaling of the $\gamma$-ray luminosity with $M$ holds even when the $\gamma \gamma$ absorption is important. The spectral distribution of Comptonized radiation changes slightly with the  black hole mass (due to the change of the synchrotron emission, which affects the position of Comptonization bumps, see  Fig 5a), however, the effect is insignificant for the $\gamma \gamma$ opacity and we have checked that it negligibly affects the observed $\gamma$-ray spectra.

\section{Discussion}

The form of the dissipation rate and the proton distribution function are two  major uncertainties for predicting the $\gamma$-luminosity. We briefly discuss here the related effects in accretion flows.

\subsection{Viscous heating}
\label{sec:diss}
The form of  the viscous dissipation rate given by equation (\ref{eq:ions}) results from the usual assumption that the viscous stress is proportional to the total pressure, with the proportionality coefficient $\alpha$. We use this form of the viscous stress for computational simplicity, however,  as we discuss in N12, this may be an oversimplified approach (although it has support in MHD simulations, see below).
Below we briefly compare the dissipation rate in our models with that predicted by the classical Novikov \& Thorne (1973) model.
We define $Q_{\rm vis}$ integrated over the whole body of the flow    as the total dissipation rate,   $Q_{\rm vis,tot}$, in our solutions  and compare it to the dissipation rate in a Keplerian disc for the corresponding value of $a$, $Q_{\rm NT,tot}$, given by the Novikov \& Thorne model. Obviously, $Q_{\rm vis,tot}$ does not need to match $Q_{\rm NT,tot}$ closely. The latter value is calculated under the assumption that the shear stress vanishes at the radius of the innermost stable circular orbit (ISCO), which condition is not applicable to geometrically thick ADAFs and its release should lead to stronger dissipation. On the other hand, ADAFs are sub-Keplerian which property decreases the dissipation rate.

Our $Q_{\rm vis,tot}$ does not differ significantly from $Q_{\rm NT,tot}$, however, the efficiency of dissipation in our model comparatively increases with increasing $a$. Specifically, $Q_{\rm vis,tot}=0.025 \dot M c^2 \approx 0.5Q_{\rm NT,tot}$ for $a=0$, $Q_{\rm vis,tot}=0.2 \dot M c^2 \approx Q_{\rm NT,tot}$ for $a=0.95$ and $Q_{\rm vis,tot}=0.58 \dot M c^2 \approx 1.5Q_{\rm NT,tot}$ for $a=0.998$. Note that in an ADAF with a large value of $a$ most of the dissipation occurs very deep in the potential, where relativistic effects strongly reduce the energy escaping to infinity, whereas in a Keplerian disc the dissipation is less centrally concentrated and its radiation is subject to less severe reduction. For example, the gravitational redshift alone (neglecting, e.g.,  the photon capture under the event horizon) would give similar luminosities received by distant observers ($\simeq 0.3 \dot M c^2$) both in our ADAF and in Novikov \& Thorne models with $a=0.998$, if all of the dissipated energy were converted into radiation. Obviously, in optically thin ADAFs only a small part of the dissipated energy is radiated away and most of it is accreted by the black hole, so the radiative efficiency is much smaller than 0.3.

As we mentioned in Section \ref{sec:flow}, the model formally allows for two solutions and the above values of $Q_{\rm vis,tot}$ correspond to the standard solution, considered in previous sections. The superhot solution has an extreme dissipation, with $Q_{\rm vis,tot} \simeq 5$--$10 \dot M c^2$. This is not necessarily an unphysical property (note that $Q_{\rm vis,tot}$ is defined in the rest frame of the flow), as the observed luminosity of such flows does not exceed the accretion rate of the rest mass energy. Nevertheless, the magnitude of $Q_{\rm vis}$ suggests that the underlying assumption of $Q_{\rm vis}\propto p$ breaks down at very large $p$.

The issue of the proper description of the $Q_{\rm vis}$ term could be partially resolved by comparing  analytic models such as our, aiming at a precise calculation of the produced radiation, with MHD simulations. The latter currently neglect radiative cooling, or use very approximate descriptions for it, in turn, they provide more accurate accounts of the dissipation physics. Such MHD simulations support some properties of our model, e.g.\ the $Q_{\rm vis}\propto p$ prescription of viscous heating (Ohsuga et al.\ 2009). Furthermore, the GR MHD simulations have shown that flows cross the ISCO without any evidence that the shear stress goes to zero, which leads to the increase of the radiative efficiency, and  the deviations from Novikov \& Thorne model increase with increasing $H/R$ ratio (e.g.\ Noble, Krolik \& Hawley 2009, Penna et al.\ 2010). We could not, however, quantitatively compare our models with such simulations, as the published results have a much smaller aspect ratio than our solutions ($H/R>0.5$).

Lastly, we remark that Gammie \& Popham (1998) and Popham \& Gammie (1998) present a model similar to ours but with a more elaborate description of the shear stress  (in turn, they neglect radiative processes). We note that their results indicate a similar in magnitude stabilizing effect of the rotation of black hole.

\subsection{Proton distribution function}
\label{sec:protons}

As pointed out by Mahadevan \& Quataert (1998), Coulomb collisions are too inefficient to thermalize protons in optically thin flows and the proton distribution function is determined by the viscous heating mechanism, which is poorly understood. Our solutions, with $M =2 \times 10^8 M_{\odot}$ and $\dot m = 0.1$, give  the accretion time-scale, $t_{\rm a}$, much shorter than the proton relaxation time-scale, $t_{\rm pp}$; e.g.\ at $r \le 20$,  $t_{\rm a} < 10$ hours and $t_{\rm pp} >  10^5$ hours. Clearly, the protons cannot redistribute their energy through Coulomb collisions.

In solar flares, the best observationally studied example of particle acceleration/heating in a magnetised plasma, a significant fraction of the released  energy is carried by non-thermal, high energy particles (e.g.\ Aschwanden 2002), which strongly motivates  for considering the nonthermal distribution of protons, as originally proposed in M97. Applying the generic description of particle acceleration, see e.g.\ section 3 in Zdziarski, Malzac \& Bednarek (2009), we check whether the conditions in ADAFs allow for proton acceleration to ultrarelativistic energies, as assumed in our computations for the power-law distributions. The magnetic field strength in our ADAF solutions is $B \simeq 10$ G, 100 G and 1000 G at $r=100$, 10 and 2, respectively. Assuming an acceleration rate ${\rm d}E/{\rm d}t \propto \xi e B$, where $e$ is the elementary charge and $\xi$ is the acceleration efficiency ($\xi \la 1$), we find that the  maximum Lorentz factor limited by  the synchrotron energy loss is  $\gamma_{\rm max} \sim 10^6-10^7$, depending on $r$. Another condition, of the Larmor radius being smaller than the acceleration site size, $R_{\rm acc}$, gives $\gamma_{\rm max} \sim 10^7$ at $r \gg 100$, and larger values of $\gamma_{\rm max}$ at smaller $r$, even if we safely assume $R_{\rm acc} = 1 R_{\rm g}$ ($ = 3 \times 10^{13}$ cm). We conclude that the central region of a hot accretion flow may be a site of the acceleration of protons to energies which easily allow hadronic emission of photons even in the  TeV range. This conclusion remains valid for the whole relevant range of accretion rates and masses of supermassive black holes, as the strength of the magnetic field scales as $B \propto \dot m^{1/2} M^{-1/2}$.

Considering processes which could compete with proton-proton interactions, we note that proton-photon interactions are much less efficient in the central region of the flow. Namely, the number density of $\gamma$-ray photons, which can effectively interact with protons in photomeson production, is by a factor of $\sim 10^3$ smaller than the number density of protons within the innermost few $R_{\rm g}$; the number density of hard X-ray photons, which can interact with protons in photopair production, is similar to $n_{\rm p}$. The cross-section for both channels of proton-photon interactions is by over two orders of magnitude smaller than the cross-section for proton-proton interaction,  making the photo-hadronic production of secondary particles negligible.

\section{Comparison with observations}
\label{sec:obs}

We briefly compare here predictions of our model with $\gamma$-ray observations of objects which may be powered by ADAFs. We note, however, that for a more detailed comparison   additional physical  processes, related to charged pion production (cf.\ Mahadevan 1999), should be taken into account. In particular, relativistic electrons produced by the pion decay should be important in modelling  the emission in the MeV range. We also tentatively discuss the origin in the accretion flow   of the very high energy radiation detected from M87 and Sgr A$^\star$, for the latter under assumption that the contributions from two separate sources dominate above 100 GeV (Sgr A$^\star$) and at lower energies (diffuse emission) in the radiation observed from the Galactic Center region. However, we note that  the opacity at very high energies may be affected by the nonthermal synchrotron emission of the relativistic electrons. The work on the model implementing the effects of charged pions is currently in progress.

\subsection{Misaligned AGNs}
\label{sec:fr1}

The misaligned AGNs detected by the {\it Fermi}-LAT include seven  FR I radio galaxies and four FR IIs (Abdo et al.\ 2010b). The low-power FR I galaxies are supposed to be powered by radiatively inefficient accretion flows (e.g.\ Balmaverde, Baldi \& Capetti 2008) and, thus, are more relevant for application of our results. Interestingly, Wu, Cao \& Wang (2011) assess that supermassive black holes in FR Is rotate rapidly, with  $a>0.9$. The X-ray emission of more luminous FR Is is supposed  to come from an accretion flow (e.g.\ Wu, Yuan \& Cao 2007), but their $\gamma$-ray emission tends to be interpreted in terms of jet emission (e.g.\ Abdo et al.\ 2009, 2010ab). FR Is are supposed to be the parent population of BL Lac objects, however, the Lorentz factors required by a jet model are much lower than typical values found in models of BL Lac objects (see, e.g., Abdo et al.\ 2010b). Therefore, the radiation observed in FR Is and BL Lacs must have a different origin, which adds some complexity to the jet model for FR Is. At least two FR I galaxies reported in Abdo et al.\ (2010b), M87 and Centaurus A, are detected because of their proximity rather than a small inclination angle of their jets and, then, they  are interesting targets for searching for the $\gamma$-ray emission from accretion flows. 

In Nied\'zwiecki et al.\ (2012a) we roughly compared preliminary results of our model with the X/$\gamma$-ray observational data of Centaurus A from {\it INTEGRAL} and {\it Fermi}-LAT. We found that the ADAF model, which matches the X-ray emission in this object, predicts the $\gamma$-ray emission significantly weaker than measured  by {\it Fermi} if the value of the spin parameter $a$ is small. We should emend, however, that this conclusion is valid only for models assuming a significant $\gamma$-ray emission only from the central $\sim 10R_{\rm g}$, such as our models T and H. Regardless of the values of $a$ and $\delta$, our model N with $s=2.6$ predicts the absorbed $L_{\rm 0.1-10 GeV}$ approximately consistent with $1.3 \times 10^{41}$ erg/s measured in Cen A by {\it Fermi}  (Abdo et al.\ 2010a). Also regardless of the value of $a$, our models with $\delta=10^{-3}$ predict the 2-10 keV flux as well as the X-ray spectral index consistent with that observed in Cen A; for $\delta \ga 0.1$ the model overpredicts the flux and hardness of the X-ray radiation. In the above we assumed the black hole mass of $2 \times 10^8 M_\odot$ (e.g.\ Marconi et al.\ 2001), which is also within the range allowed by the recent measurement of Gnerucci et al.\ (2011).

The  central  core of the accretion system in M87 has been considered as the $\gamma$-ray emitting region  e.g.\ by Neronov  \&  Aharonian  (2007) in their model with particle acceleration in the black hole magnetosphere. The accretion rate assessed from the high-resolution observations of the nucleus of M87 by {\it Chandra}, and used to model the multiwavelength spectrum of the nucleus by emission from an ADAF, by Di Matteo et al.\ (2003; note that their definition of $\dot m$ differs from ours by a factor of 10),      $\simeq 0.1 M_\odot$/year, corresponds to $\dot m \simeq 0.01$ for a black hole with $M= 3 \times 10^9 M_\odot$. At such $\dot m$ the central region should be transparent to $\gamma$-ray photons. Then, we compare the unabsorbed luminosities from our nonthermal models with $s=2.2$ (approximately consistent with the slope of the {\it Fermi} data above 200 MeV; Abdo et al.\ 2009) and  $M= 3 \times 10^9 M_\odot$ with the $\gamma$-ray measurements of M87. We find that the luminosity derived in the 0.2-10 GeV range by {\it Fermi} (Abdo et al.\ 2009), and above 100 GeV by ground-based telescopes (e.g.\ Aleksi\'c et al.\ 2012) in the low state of M87, can be reproduced by our model with $a = 0.998$ and $\simeq 0.14 M_\odot$/year (for $\theta_{\rm obs} = 40\degr$). The required accretion rate, larger by 40 per cent than the face value of the estimate in Di Matteo et al.\ (2003), is allowed by the precision of the estimation of the accretion rate  using Bondi accretion theory. We conclude that hadronic processes in ADAF can contribute significantly to the $\gamma$-ray emission observed in M87 in the low state, or even explain these observations  entirely. However, such a model requires that all the available material forms the innermost flow, i.e.\ it does not allow a strong
reduction the accretion rate in the central region by outflows (assumed in models of Sgr A$^\star$, see below).

\subsection{Sgr A$^\star$}
\label{sec:astar}

The value of $\dot m$ in the central region is the major issue for applications of ADAF models to the supermassive black hole in the Galactic Center. The Bondi accretion rate  corresponds to $\dot m_{\rm B} \simeq 10^{-3}$ for $M \simeq 3 \times 10^6 M_\odot$ and early  models used such $\dot m$ to explain the broad-band spectra of Sgr A$^\star$ (e.g.\ Narayan et al.\ 1998). The measurement of the millimetre/submillimetre polarization of Sgr A$^\star$ is often assumed to limit the accretion rate to $\dot m \le 3 \times 10^{-5}$ (e.g.\ Marrone et al.\ 2007), but counterarguments are presented, e.g., by Mo\'scibrodzka, Das \& Czerny (2006) and Ballantyne, \"Ozel \& Psaltis (2007). An updated ADAF model to Sgr A$^\star$, with $\dot m \simeq 10^{-5}$ and strong electron heating,  is presented in Yuan, Quataert \& Narayan (2003). An alternative to ADAF model, proposed by Mo\'scibrodzka et al.\ (2006) and recently applied by Okuda \&  Molteni (2012), with the low angular momentum flow, assumes $\dot m \simeq 6 \times 10^{-4}$. 
The low angular momentum flow has a similar density in the innermost region as an ADAF with the same $\dot m$, but much smaller proton temperature $< 4 \times 10^{11}$ K. Then, the magnitude of hadronic processes could be used to distinguish the two classes of models; however, the low $T$ may be an artificial effect resulting from the neglect of viscosity in the former (low angular momentum) class.

The {\it CGRO}/EGRET source, 3EG J1746-2851, was initially considered as a possible $\gamma$-ray   counterpart of  Sgr A$^\star$; Narayan et al.\ (1998) and OM03  found that their ADAF models underpredicted the  $\gamma$-ray luminosity implied by the EGRET measurement. However, improved analyses (e.g.\ Pohl 2005) subsequently indicated that 3EG J1746-2851 is displaced from the exact Galactic Center and excluded Sgr A$^\star$, as well as the TeV source observed with HESS which may be directly related with Sgr A$^\star$, as its possible counterparts.

Recently, Chernyakova et al.\ (2011) analysed the {\it Fermi}-LAT observations of the Galactic Center and  combined them with the HESS observational data of the point-like source (Aharonian et al.\ 2009). The spectrum of the source seen in the MeV/GeV band by {\it Fermi} is consistent with a $\pi^0$-decay spectrum.  The HESS data indicate flattening of the spectrum above 100 GeV, suggesting that at the highest energies a different spectral component dominates, with rather hard spectrum, $\Gamma \simeq 2.2$. In the $EF_E$ plot, the normalization of the HESS source is an order of magnitude larger than the quiescent X-ray emission of Sgr A$^\star$. To explain these observations, Chernyakova et al.\ (2011) discuss a model, following previous works (e.g.\ Atoyan \& Dermer 2004), with protons accelerated in the accretion flow and then diffusing outwards to interact with dense gas at  distances of $\sim 1$ pc. The emission at energies below 100 GeV may be explained by  interactions of protons injected into the interstellar medium during a strong flare of   Sgr A$^\star$ that occurred 300 years ago, as protons generating photons with such energies are still diffusively trapped in the $\gamma$-ray production region. On the other hand,  most of the higher energy protons have already escaped and hence  an additional, persistent injection of high-energy protons has to be assumed to account for the HESS observations, with the required rate of $2 \times 10^{39}$ erg/s implying a very high efficiency of the conversion of the accreting rest mass energy into the  proton energy.

We remark that alternatively the spectral component above 100 GeV can be explained by proton-proton interactions in an accretion flow around a rapidly rotating black hole. Our model with $a=0.998$, $s=2.2$ and edge-on viewing direction predict the flux consistent with the HESS detection for $\dot m = 3 \times 10^{-4}$.  The scenario with the $\gamma$-ray  emission produced in the accretion flow may be tested over the following years, if the fall of the gas cloud into the accretion zone of Sgr A$^\star$ (Gillessen et al. 2012) results in the increase of the mass accretion rate.

\subsection{Seyfert galaxies}
\label{sec:sy}

Spectral properties of the Seyfert galaxy, NGC 4151, are consistent with the model of an inner hot flow surrounded by an outer cold disc (e.g.\ Lubi\'nski et al.\ 2010), however, its rather large luminosity, $L > 0.01 L_{\rm Edd}$ indicates that a potential $\gamma$-ray emission from innermost region would be strongly absorbed. The constraint of $L_{\rm 0.1-10 GeV}/L_{\rm 14-195 keV} < 0.0025$ derived for this object in Ackermann et al.\ (2012) can still give interesting information if the spectral model constrains the parameters of the inner hot/outer cold accretion system (in particular, the distance of transition between the two modes of accretion). The constraint of $L_{\rm 0.1-10 GeV}/L_{\rm 14-195 keV}<0.1$, or even $<0.01$ for some objects, found for other Seyfert galaxies by Ackermann et al.\ (2012)  is not strongly constraining for the spin value, as can be seen in Fig.\ \ref{fig:9} (note that $L_{\rm 14-195 keV}$ is by a factor of a few, depending on $\Gamma_X$, larger than $L_{\rm 2-10 keV}$), especially for large X-ray luminosities (and hence strong $\gamma \gamma$ absorption) characterising most of objects analysed in that paper.

\section{Summary}

We have studied the $\gamma$-ray emission resulting from proton-proton interactions in two-temperature ADAFs. Our model relies on the global solutions of the GR hydrodynamical model, same as  OM03, but we improve   their computations by taking into account the relativistic  transfer effects as well as the $\gamma \gamma$ absorption and by properly describing the global Comptonization process.

We have found that the spin value is reflected in the properties of $\gamma$-ray emission, but the effect is not thrilling. The speed of the black hole rotation strongly affects the $\gamma$-ray emission produced within the innermost $10 R_{\rm g}$. If emission from that region dominates, the observed $\gamma$-ray radiation depends on the spin parameter noticeably; the intrinsic $\gamma$-ray luminosities of flows around rapidly-rotating and non-rotating black holes  differ by over a factor of $\sim 10$. However, if the $\gamma$-ray emitting region extends to larger distances, the dependence is reduced. In the most extreme case with protons efficiently accelerated to relativistic energies in the whole body of the flow, the regions within and beyond $10 R_{\rm g}$ give comparable contributions to the total emission, reducing the difference of $\gamma$-ray luminosities between high and low values of $a$ to only  a factor of $\sim 2$. The radial emissivity profile of $\gamma$-rays is very uncertain (as the acceleration efficiency may change with radius), then, the level of the $\gamma$-ray luminosity cannot be regarded as a very sensitive probe of the spin value. Still, it may be possible to assess a slow rotation in a low luminosity object by putting an upper limit on the $\gamma$-ray luminosity at the level of $\la 0.01$ of the X-ray luminosity. The presence of nonthermal protons may be easily assessed from the $\gamma$-ray spectrum as for a purely thermal plasma the total $\gamma$-ray emission would be observed only at energies lower than 1 GeV.

We have considered  the accretion rate of $\dot m = 0.1$, for which our model gives the bolometric luminosities between $\simeq 4 \times 10^{-4} L_{\rm Edd}$ and $10^{-2} L_{\rm Edd}$. Such  $\dot m$, with the corresponding range of $L/L_{\rm Edd}$, seems to be  favoured for investigation of the hadronic $\gamma$-ray emission, with the related effects of the space-time metric, because the internal $\gamma$-ray emission is large and its attenuation by $\gamma \gamma$ absorption is weak. Flows with $L > 10^{-2} L_{\rm Edd}$ can produce observable $\gamma$-ray radiation only if the emitting region extends out to a rather large distance of several tens of $R_{\rm g}$. 

We have found the  X-ray to $\gamma$-ray luminosities ratio  as a function of the black hole spin and the efficiency of the direct heating of electrons. The  $L_{\rm 0.1-10 GeV}/L_{\rm 2-10 keV}$ ratios reaching $\sim 0.1$, with the corresponding levels of the $\gamma$-ray luminosities which may be probed in nearby AGNs at the current sensitivity of {\it Fermi}-LAT surveys, encourage to consider contribution from an accretion flow to the $\gamma$-ray emission observed in low-luminosity objects. We point out that such a contribution should be strong at least in Cen A.

The $\gamma$-ray luminosity decreases rapidly ($\propto \dot m^2$) with decreasing $\dot m$. M87 and Sgr A$^\star$ are the obvious, however possibly also unique, objects  for which the $\gamma$-ray emission from the flow can be searched at $\dot m \ll 0.1$. 
Obviously, contribution from other $\gamma$-ray emitting sites should be properly subtracted to establish the luminosity of an accretion flow, which may be particularly difficult in  Sgr A$^\star$. The ADAF model with a hard (acceleration index $s \simeq 2.2$) nonthermal proton distribution can explain the $\gamma$-ray detections of both Sgr A$^\star$ (above 100 GeV) and M87, however, it does not allow for strong reduction of the accretion rate by outflows. Nevertheless, it seems intriguing that in both nearby, low accretion-rate objects, in which the $\gamma$-ray radiation is not suppressed by $\gamma \gamma$ absorption, observations reveal very high energy components, consistent with predictions of such a model. In both objects the model requires a large black hole spin.

The luminosity ratio should decrease rather slowly ($L_{\gamma}/L_{\rm X}\propto \dot m^{0.3}$, as the radiative efficiency of hot flows varies as $\dot m^{0.7}$ at small $\dot m$; cf.\ Xie \& Yuan 2012) with decreasing $\dot m$. The presence of the $\gamma$-ray signal in low-luminosity AGNs is often considered as the evidence for the origin of their radiation in a jet (see, e.g., Takami 2011).   We note that such a diagnostic is not valid because accretion flows may produce a similarly strong $\gamma$-ray emission as jets in such objects.

\section*{ACKNOWLEDGMENTS}

We are grateful to W\l odek Bednarek for helpful discussions and to the referee for a very careful review and comments which helped us to improve the presentation of the results. This research has been supported in part by the Polish NCN grant N N203 582240. FGX has been supported in part by the NSFC (grants 11103059, 11121062, 11133005, and 11203057), the NBRPC (973 Program 2009CB824800), and SHAO key project No.\ ZDB201204.

\label{lastpage} 
\end{document}